\documentclass[12pt]{article}

\usepackage[textheight=22.5truecm,textwidth=16.8truecm,top=2.95truecm,left=2.15truecm]{geometry}
\usepackage{
amsmath,
amsthm,
amssymb,
ascmac,
bm,
tikz,
braket,
mathrsfs,
graphicx,
color,
hyperref,
cite,
titlesec,
authblk,
caption,
enumitem
}

\hypersetup{colorlinks,bookmarksopen,bookmarksnumbered,citecolor=black,linkcolor=black,linktocpage,pdfstartview=FitH,urlcolor=black}

\numberwithin{equation}{section}

{\makeatletter
\long\def\@makefntext#1{\parindent 1em\noindent 
\@hangfrom{\hbox to 1.8em{\hss$^{\@thefnmark}$}}#1}
\makeatother}
\renewcommand\thefootnote{\arabic{footnote})}

\def\fnum@figure{\textbf{\figurename\nobreakspace\thefigure}}
\def\fnum@table{\textbf{\tablename\nobreakspace\thetable}}
\long\def\@makecaption#1#2{%
  \vskip\abovecaptionskip
  \sbox\@tempboxa{\small #1. #2}%
  \ifdim \wd\@tempboxa >\hsize
    \small #1. #2\par
  \else
    \global \@minipagefalse
    \hb@xt@\hsize{\hfil\box\@tempboxa\hfil}%
  \fi
  \vskip\belowcaptionskip}
\renewcommand{\l}[0]{\left}
\renewcommand{\r}[0]{\right}
\renewcommand{\d}[0]{\mathrm{d}}

\newcommand{\ketbra}[2]{\ket{#1} \hspace{-4.5pt} \bra{#2}}
\newcommand{\reft}[0]{{\mathrm{ref},t}}
\newcommand{\red}[1]{#1}

\allowdisplaybreaks

\title{\hfill\parbox{3cm}{
\normalsize
KUNS-2995
}\\[12pt]
Coarse-graining black holes out of equilibrium\\ with boundary observables on time slice
}

\author[1]{Daichi Takeda}

\affil[1]{\it Department of Physics, Kyoto University, Kyoto 606-8502, Japan}

\date{}

\begin{document}
\maketitle

\renewcommand{\thefootnote}{\fnsymbol{footnote}}
\footnote[0]{
takedai@gauge.scphys.kyoto-u.ac.jp
}
\renewcommand{\thefootnote}{\arabic{footnote}}

\begin{abstract}
In black hole thermodynamics, defining coarse-grained entropy for dynamical black holes has long been a challenge, and various proposals, such as generalized entropy, have been explored.
Guided by the AdS/CFT, we introduce a new definition of coarse-grained entropy for a dynamical black hole in Lorentzian Einstein gravity.
On each time slice, this entropy is defined as the horizon area of an auxiliary Euclidean black hole that shares the same mass, (angular) momenta, and asymptotic normalizable matter modes with the original Lorentzian solution.
The entropy is shown to satisfy a generalized first law and, through holography, the second law as well.
Furthermore, by applying this thermodynamics to several Vaidya models in AdS and flat spacetime, we discover a connection between the second law and the null energy condition.
\end{abstract}

\newpage
\tableofcontents

\newpage
\section{Introduction and summary}
Gravity is considered thermodynamic \cite{BardeenCarterHawking}.
The zeroth law asserts the existence of intensive variables, which in black hole thermodynamics correspond to quantities constant over the event horizon, such as surface gravity and angular velocity.
The first law is one of the key concept, stating the energy conservation among neighboring stationary black holes \cite{HawkingHartle1st, Wald:1993nt} when the horizon area is viewed as the entropy \cite{Bekenstein1}.
Applied to the local Rindler patch, the first law reproduces the Einstein equation, allowing us to regard it as an equation of state \cite{Jacobson:1995ab}.
Above all, not only does black hole thermodynamics apply as a formal analogy, but it also has  physical reality in the Hawking radiation \cite{HawkingRad}.

Thermodynamics helps us understand spacetime physics, because it gives a macroscopic constraints that the statistical mechanics of a quantum theory of gravity must adhere to.
First, the microscopic counting of states in string theory (beginning with \cite{Strominger:1996sh}) agrees with the horizon area, confirming the importance of exploring fundamental theories through macroscopic observations.
Second, given the entropy as a function of sufficient extensive variables, thermodynamics dictates among what states transitions are allowed to happen.
This is the role of the second law.
Since general relativity itself does not teach what ``physical" time evolutions are, energy conditions are necessary for describing ``physical" systems.
Here, the thermodynamics of spacetime is expected to provide an answer if established.

However, the second law remains under debate.
While the Hawking area theorem \cite{Hawking2nd} guarantees that the horizon area does not decrease toward the future, the question as to whether the horizon area can be regarded as the entropy even out of equilibrium is nontrivial.
Besides, it is incompatible with the Hawking radiation occurring in the semiclassical regime.
An alternative proposal that has been extensively studied is the \textit{generalized entropy} \cite{PhysRevD.7.2333, PhysRevD.9.3292}, which is the sum of the entropy of matter outside the horizon on a time slice and the area of the cross section between the horizon and the time slice.
Despite numerous attempts, however, a proof of the \textit{generalized second law} valid in any situation, or an agreement on a suitable definition of the entropy for dynamical black holes has not been achieved (see \cite{Wall:2009wm, Carlip:2014pma, Wall:2018ydq, Sarkar:2019xfd} for review).
The second law is one of the primary topics addressed in this paper.\footnote{
The third law seems not to hold in general for black hole physics \cite{PhysRevD.56.6467, Kehle:2022uvc}, but it is not a drawback since the third law is just a phenomenological observation in laboratories, and in principle, it is not necessary in the axiomatic construction of thermodynamics.
}
\vspace{12pt}

Gravity, on the other hand, is considered holographic.
In the classical Einstein gravity, the bulk contribution of the on-shell action vanishes and meaningful values arise from the boundary term, by which we define the energy, (angular) momenta, and other charges.
At the same time, the entropy in stationary cases is defined on the causal boundary, the horizon, and interestingly, the first law holds among those boundary values \cite{Wald:1993nt}.

The holographic nature is a key concept also at the quantum level \cite{tHooft:1993dmi, Susskind:1994vu, Bousso:1999dw, Bousso:2002ju}, having provided hints towards quantum gravity.
Among those, the AdS/CFT correspondence \cite{Maldacena:1997re,Witten:1998qj,Gubser:1998bc} has attracted significant attention so far.
According to the correspondence, the gravitational degrees of freedom are mapped onto the quantum field theory defined on the AdS boundary.
Notably, the entanglement entropy is expected to reveal the nature of the dual spacetime \cite{Ryu:2006bv, Ryu:2006ef, Hubeny:2007xt, Wall:2012uf}.
The entanglement entropy of a boundary region is equivalent to the area of an extremal surface in the bulk, and in the semiclassical regime the von Neumann entropy of matter is added to it as quantum corrections \cite{Barrella:2013wja, Faulkner:2013ana, Engelhardt:2014gca}; the total entropy takes a form similar to the thermodynamic generalized entropy explained above.
This direction recently developed into the \textit{Island formula} \cite{Penington:2019npb, Almheiri:2019psf, Almheiri:2019hni, Penington:2019kki, Almheiri:2019qdq} to study the Page curve \cite{Page:1993wv, Page:2013dx},  with applications extending beyond AdS gravity \cite{Hashimoto:2020cas}.

When applying holography to black hole thermodynamics, particularly concerning the second law, it seems better to introduce coarse-grained entropies. 
This is because, for instance, the unitarity of the CFT ensures that the fine-grained entropy of a total system state is time-independent, which makes the HRT surface \cite{Hubeny:2007xt} insensitive to any probe sent from the boundary \cite{Takayanagi:2010wp}.
Within the framework of the AdS/CFT, a coarse-grained entropy associated with the second law in the bulk was first explored in \cite{Kelly:2013aja}, where the \textit{one-point entropy} was introduced.
The relationship with the first law was also investigated in \cite{Kelly:2014owa}.
In \cite{Kelly:2013aja}, the bulk dual of the one-point entropy was first conjectured to be the \textit{causal holographic information} \cite{Hubeny:2012wa, Freivogel:2013zta}, but unfortunately, counterexamples were identified in \cite{Engelhardt:2017wgc}, leading the authors to conclude that the bulk counterpart seems more intricate than initially proposed.
Other approaches, as seen in \cite{Engelhardt:2017aux, Grado-White:2017nhs, Engelhardt:2018kcs, Chandra:2022fwi}, begin with defining their coarse-graining procedures  in the bulk.
While \cite{Engelhardt:2017aux, Engelhardt:2018kcs} and \cite{Chandra:2022fwi} employ different ways of coarse-graining, both approaches are associated with the apparent horizon, presenting intriguing implications to gravity and its thermodynamics.
However, as noted by the authors, none of those approaches has yet acquired a complete description within the boundary language.

\vspace{12pt}
In the same spirit, we propose via the AdS/CFT a new coarse-grained entropy, which is defined both on the boundary and in the bulk.
It is valuable to study variety of ways of coarse-graining, as the coarse-grained entropy differs depending on what information one aims to preserve, and different coarse-graining methods will offer different perspectives on gravity.
The entropy we propose satisfies a generalized version of the first law, and by the AdS/CFT the second law as well.
In addition, it coincides with the Bekenstein-Hawking entropy for stationary solutions.

We begin with the positivity of the relative entropy in the boundary quantum theory.\footnote{
The usage of relative entropy in the AdS/CFT to survey the bulk is also seen in \cite{Blanco:2013joa, Banerjee:2014oaa, Banerjee:2014ozp, Lashkari:2014kda}.
}
Let $\rho_1$ and $\rho_2$ be normalized density operators, then the relative entropy is defined as
\begin{align}
  S(\rho_1||\rho_2):=\mathrm{Tr}(\rho_1\ln \rho_1 - \rho_1\ln \rho_2),\label{eq: def of relative entropy}
\end{align}
which is always non-negative.
Suppose the initial state is a steady state $\rho_0$, which evolves to $\rho_t$ with a time-dependent Hamiltonian.
The time-dependence is triggered by sources coupling to composite operators.
On the other hand, we prepare a reference state $\rho_\reft$ sharing with $\rho_t$ the same expectation values of composite operators $\{O_I\}$ of our interest: $\mathrm{Tr}(\rho_\reft O_I) = \mathrm{Tr}(\rho_t O_I)$.
At $t = 0$ we set the reference state as $\rho_{\mathrm{ref},0} = \rho_0$.
Then, we define our coarse-grained entropy by $S_t := -\mathrm{Tr}(\rho_\reft\ln \rho_\reft)$, and will show that $S_t\geq S_0$ holds for any $t$, based on the positivity $S(\rho_t|| \rho_\reft)\geq 0$.

This coarse-grained entropy is in the bulk equivalent to the horizon area of an auxiliary Euclidean spacetime\footnote{
As is also the case in \cite{Engelhardt:2018kcs}, an appropriate auxiliary spacetime different from the original is necessary, since the coarse-graining process changes the state on the boundary.}
realized as the dominant saddle point of the gravitational path integral dual to $\rho_\reft$.
In the bulk language, $\mathrm{Tr}(\rho_\reft O_I) = \mathrm{Tr}(\rho_t O_I)$ means that the Euclidean solution shares the same asymptotic normalizable modes with the original Lorentzian time slice.
Besides, $S_t \geq S_0$ means that the horizon area of the Euclidean black hole we refer to at $t$ is always greater than or equal to the horizon area of the initial black hole.
This is the second law (for adiabatic process) in this paper.
As explained below, this is not an assertion of monotonically increasing area as in Hawking's area law.
After obtaining the gravitational description, we will also derive the first law, which, in addition to the usual terms, contains contributions from the asymptotic matter modes.

We will also check the two laws explicitly in several null-ray collapse models and find that our second law implies the null energy condition --- derived by the AdS/CFT, this can be viewed as a consequence of quantum gravity.
The gravitational description we finally obtain can formally be applied to non-AdS spacetimes.
To test its applicability, an asymptotically flat collapse is chosen as one of the examples, and the two laws actually hold in this example.

It is worth mentioning that the second law in this paper, though not monotonic, is compatible with the expressions of the second law in the context of non-equilibrium thermodynamics.
The second law in thermodynamics serves as the criterion for possible transitions between steady states.
In non-equilibrium thermodynamics, the second law is generalized to the statement that the entropy production\footnote{
The Shannon entropy change plus the heat absorption divided by temperature.
} never becomes negative.
When the total system is deterministic, the second law follows from the result of \cite{jarzynski1997nonequilibrium, jarzynski2000hamiltonian}, where the initial state is prepared as a product state of the system and bath, with the bath in equilibrium.
Since the initial state is prepared specially, the monotonicity is not guaranteed.
The second law remains monotonic, for example, when the system is Markovian \cite{evans1993probability, kurchan1998fluctuation}, as a Markovian system does not care about its history, making the initial state not special anymore.
It may not be obvious whether the monotonicity holds for dynamical gravitational systems as well.
The quantum versions of the second law without the monotonicity were also shown via the relative entropy in \cite{tasaki2000jarzynski, esposito2010entropy} (see \cite{Sagawa:2012eqh} for review).
The organization of this paper is as follows.
In section \ref{sec: CFT}, the second law is derived in the QFT.
In section \ref{sec: gravity}, we rewrite it in the bulk Einstein gravity, and also derive the (generalized) first law.
In section \ref{sec: demo}, the derived thermodynamic laws are applied to Vaidya-type models, indicating that thermodynamics constraints gravity.

\subsubsection*{Summary of the results}
Here, the main results in the gravity side is roughly overviewed.
Let $M$ be a $(d+1)$-dimensional Lorentzian manifold with timelike boundary $\partial M$.
The theory is supposed to be the Einstein theory with matter fields, whose action is denoted as  $I_{\mathrm{grav}}$ (see \eqref{eq: gravity action}).
Gauge fields like Maxwell field can be included.

We suppose that a stationary configuration is realized on the initial time slice.
This is prepared by analytically continuing a Euclidean solution (figure \ref{fig: Lorentzian bulk}).
To define a coarse-grained entropy at time $t>0$, we need to specify a set of fields to be respected, whose values on $\partial M$ we write as $\{w^I_t(\theta)\}$.
Here, $I$ is the label of fields, and the boundary coordinate is written as $z = (t,\theta^A)$ with $A = 1,\cdots,d-1$.
The induced metric on $\partial M$ is assumed static, while each $w^I_t(\theta)$ can depend on time.
The normalizable mode conjugate to $w^I_t(\theta)$ is defined to be
\begin{align}
  \pi_{I,t}(\theta) :=  \frac{\delta}{\delta w^I_t(\theta)}\left(\left. I_{\mathrm{grav}} \right|_{\mathrm{on-shell}}[w] \right).
\end{align}
We write the mass and the (angular) momenta in $\theta^A$-direction as $h_t$ and $p_{A,t}$, both of which are defined by the Brown-York tensor \eqref{eq: BY tensor}.
Note that $\pi_{I,t}(\theta)$ is a local quantity, dependent on the spatial coordinates $\theta^A$, while $h_t$ and $p_{A,t}$ are independent of $\theta^A$.

Our coarse-grained entropy $S_t$ is defined by the following procedure.
On each time slice, we find the auxiliary Euclidean solution which dominates the gravitational path integral, while respecting $h_t$, $p_{A,t}$ and $\pi_{I,t}$ as its mass, (angular) momenta, and asymptotic modes.
We assume that this Euclidean solution is realized stationary, independent of the imaginary time.
Then, $S_t$ is defined to be the horizon area, the area of the cigar tip in figure \ref{fig: cigar}.
Since the initial configuration of $M$ is stationary, $S_0$ is exactly the Bekenstein-Hawking entropy of the initial black hole.

The main claims of this paper are as follows.
First, $\dot S_t$ (the dot means the time-derivative) is given by the first law, with additional contributions consisting of $\{w^I\}$ and $\{\pi_I\}$.
Second, $S_t$ satisfies the second law, $S_t \geq S_0$.
While the first law is derived within the Einstein gravity, the second law is derived via the holographic dictionary.
Nevertheless, we conjecture that it must hold also in generic cases under the setup mentioned in the beginning (see section \ref{subsubsec: non-AdS} and \ref{subsec: RN}).

\section{Coarse-grained state in QFT}\label{sec: CFT}
In this section, we first introduce the reference state $\rho_\reft$, a coarse-grained state to be compared with the original state at time $t$ through the relative entropy.
The second law is derived from the positivity of the relative entropy.
After that, in preparation for section \ref{sec: gravity}, we move on to the path integral representation.

\subsection{Coarse-grained state and second law}
\subsubsection{The simplest case}
Let us grasp the essence with a warm-up example.
We consider a generic quantum theory.
The initial state at $t=0$ is supposed to be
\begin{align}
  \rho_0 = \frac{ e^{-\beta H(0)}}{Z_0(\beta)},\qquad
  Z_0(\beta) := \mathrm{Tr} e^{-\beta H(0)},
\end{align}
 where $H(0)$ is the Hamiltonian at $t=0$ and $\beta_0$ is an arbitrary inverse temperature.
 The system evolves with a time-dependent Hamiltonian $H(t)$ as
 \begin{align}
  \rho_t = U(t)\rho_0 U(t)^{-1},\qquad
  U(t) := \mathrm{T}\exp\left(-i\int_0^t \d s\,H(s) \right).
\end{align}
We choose the following as the reference state:
\begin{align}
  \rho_\reft = \frac{ e^{-\beta' H(t)}}{Z_t(\beta')},\qquad
  Z_t(\beta') := \mathrm{Tr} e^{-\beta' H(t)},\label{eq: warm-up ansatz}
\end{align}
with $\beta'$ being any positive parameter.

In this setup, we compute the relative entropy $S(\rho_t|| \rho_\reft)$ defined in \eqref{eq: def of relative entropy}.
First, since the evolution is unitary, we have
\begin{align}
  \mathrm{Tr}(\rho_t \ln \rho_t) = \mathrm{Tr}(\rho_0 \ln \rho_0) = -\ln Z_0(\beta) - \beta \mathrm{Tr}\left(\rho_0 H(0) \right).
\end{align}
The remaining piece in \eqref{eq: def of relative entropy} is readily
\begin{align}
   \mathrm{Tr}(\rho_t \ln \rho_\reft) = -\ln Z_t (\beta') - \beta' \mathrm{Tr}\left(\rho_t H(t) \right).
\end{align}
The inequality $S(\rho_t|| \rho_\reft) \geq 0$ reads
\begin{align}
   -\mathrm{Tr}(\rho_t \ln \rho_\reft)= \beta' \mathrm{Tr}\left(\rho_t H(t) \right) +\ln Z_t (\beta') \geq  \beta \mathrm{Tr}\left(\rho_0 H(0) \right) +\ln Z_0(\beta).
  \label{eq: warm-up ineq}
\end{align}

When $\beta'$ is set to be $\beta$, the inequality gives \cite{tasaki2000jarzynski}
\begin{align}
  \mathrm{Tr}\left(\rho_t H(t) \right) - \mathrm{Tr}\left(\rho_0 H(0) \right) \geq [-\beta^{-1} \ln Z_t (\beta)] - [-\beta^{-1} \ln Z_0(\beta)].
\end{align}
The l.h.s.\ is regarded as the work by the time-dependent part of the Hamiltonian, not as the heat because the system is unitary.
The r.h.s.\ can be seen as the free energy difference.
This coincides with the maximum work principle.

In this paper, we would rather like to ask what $\beta'$ makes \eqref{eq: warm-up ineq} the tightest bound.
The condition that extremizes the l.h.s.\ of \eqref{eq: warm-up ineq} is found to be
\begin{align}
  \mathrm{Tr}\left(\rho_t H(t) \right) = \mathrm{Tr}\l(\rho_\reft H(t)\r),
  \label{eq: warm-up minimize}
\end{align}
where $\beta'$-dependence resides in $\rho_\reft$.
With $\beta'$ satisfying this, \eqref{eq: warm-up ineq} is reduced to
\begin{align}
  S(\rho_\reft) \geq S(\rho_0),\qquad S(\rho) := -\mathrm{Tr}\left(\rho\ln \rho \right).
\end{align}

Here are some comments.
First, the extremal condition \eqref{eq: warm-up minimize} actually gives the minimum, since the  l.h.s.\ of \eqref{eq: warm-up ineq} goes to positive infinity as $\beta' \to \infty$.
\red{Second, our ansatz \eqref{eq: warm-up ansatz} is justified by considering coarse-graining $\rho_t$ as follows.
When coarse-graining a state $\rho_t$ with a given operator set $\{O_I\}$ respected, we find the \textit{maximum} of $S(\rho')$ under the following condition:
\begin{align}
  \mathrm{Tr}\left(\rho' O_I \right) = \mathrm{Tr}\left(\rho_t O_I \right)\qquad \mbox{for all $O_I$ in $\{O_I\}$}.
\end{align}
This problem is solved by the optimizing
\begin{align}
  \tilde S(\rho') = S(\rho') - \lambda^I \left[\mathrm{Tr}\left(\rho' O_I \right) - \mathrm{Tr}\left(\rho O_I \right) \right] + \tilde \lambda (\mathrm{Tr}\rho -1),
\end{align}
where $\lambda^I$ and $\tilde \lambda$ are the Lagrange multipliers, and summation is taken for the label $I$.
Then the solution is found to be $\rho' = \rho_{\mathrm{cg},t} \propto \exp(-\lambda^I O_I)$.
When $\{O_I\} = \{H(t)\}$, the coarse-grained state $\rho_{\mathrm{cg},t}$ is exactly \eqref{eq: warm-up ansatz}, so $S(\rho_{\mathrm{ref},t})$ is viewed as the coarse-grained entropy.
Since the coarse-graining is a process of treating the system thermodynamically, the ansatz \eqref{eq: warm-up ansatz} is appropriate.
}

\subsubsection{Generalization}
We consider more general situations in a relativistic quantum field theory on $d$-dimensional spacetime, whose metric is supposed to be static.
The coordinate is written as $z^a = (t, \theta^A)$ with $t$ being the time coordinate, and the metric as
\begin{align}
  \d s^2 = \sigma_{ab}\d z^a \d z^b = -\d t^2 + \sigma_{AB}(\theta) \d\theta^A \d\theta^B\qquad
  (\sigma := \mathrm{det}\, \sigma_{ab} = -\mathrm{det}\, \sigma_{AB}).
  \label{eq: boundary metric}
\end{align}
The action is supposed to take the form
\begin{align}
  I_{\mathrm{QFT}}[w] = \int\d^d z \sqrt{-\sigma} \l(\mathcal{L} + w^I(z) O_I(z)\r),
  \label{eq: CFT Lagrangian}
\end{align}
where $\mathcal{L}$ is the Lagrangian density without explicit time-dependence, each $O_I(z)$ is a composite operator (without time-derivative), and $w^I(z)$ is a source coupling to it.
The Hamiltonian operator is then given by
\begin{align}
  H(t) = H - \int \d ^{d-1}\theta \sqrt{-\sigma}\, w^I(t,\theta) O_I(\theta),
  \label{eq: H(t)}
\end{align}
where $H$ is the Hamiltonian when $w = 0$.
In this expression and below, we adopt the Schr\"odinger picture.
Let $T_{ab}(\theta)$ be the stress tensor operator when $w = 0$.
Then, the Hamiltonian $H$ and momentum $P_A$ without sources are given as
\begin{align}
  H = \int \d ^{d-1}\theta \sqrt{-\sigma} \, u_a T^a_b t^b,\qquad
  P_A = \int \d ^{d-1}\theta \sqrt{-\sigma} \, u_a T^a_b e^b_A,\label{eq: H and P}
\end{align}
where $u_a$ is the unit normal to time slices, $t^a\partial_a = \partial_t$, and $e^b_A\partial_b = \partial_A$.
For the metric \eqref{eq: boundary metric}, we have $u^a = t^a$.

At $t=0$, we prepare the initial state as
\begin{align}
  \rho_0 = \frac{1}{Z_0} \exp\left[-\beta \left(H - \omega^A P_A - \int \d ^{d-1}\theta \sqrt{-\sigma} \, j^I(\theta) O_I(\theta) \right) \right],\label{eq: rho_0}
\end{align}
with $Z_0$ determined through $\mathrm{Tr}\rho_0 = 1$ and dependent on $\beta_0, \omega^A_0$ and $j^I_0$.
This state evolves according to \eqref{eq: H(t)}:
\begin{align}
  \rho_t = U(t)\rho_0 U(t)^{-1},\qquad
  U(t) := \mathrm{T}\exp\left(-i\int_0^t \d s\,H(s) \right).\label{eq: rho_t}
\end{align}
We choose the reference state as
\begin{align}
  \rho_\reft = \frac{1}{Z_t} \exp\left[-\beta_t \left(H - \omega^A_t P_A - \int \d ^{d-1}\theta \, \sqrt{-\sigma} j^I_t(\theta) O_I(\theta) \right) \right].
  \label{eq: rho_ref}
\end{align}
The parameters $\beta_t$, $\omega_t^A$, and $j_t^I(\theta)$ are the Lagrange multipliers to be optimized below, and as the initial conditions, we require
\begin{align}
  \beta_{t=0} = \beta,\qquad
  \omega^A_{t=0} = \omega^A,\qquad
  j^I_{t=0} = j^I.\label{eq: initial cond}
\end{align}

The relative entropy between $\rho_t$ and $\rho_\reft$ is calculated as
\begin{align}
  S(\rho_t|| \rho_\reft) 
  = - S(\rho_t) + \mathrm{Tr}\left(\rho_t \ln \rho_\reft \right)
  = - S(\rho_0) + \mathrm{Tr}\left(\rho_t \ln \rho_\reft \right) \geq 0.\label{eq: ineq}
\end{align}
The tightest bound is achieved by optimizing this with respect to $(\beta_t, \omega_t, j_t)$, and the conditions are found to be
\begin{align}
  \mathrm{Tr}\left(\rho_t H \right) = \mathrm{Tr}\left(\rho_\reft H \right),~~
  \mathrm{Tr}\left(\rho_t P_A \right) = \mathrm{Tr}\left(\rho_\reft P_A \right),~~
  \mathrm{Tr}\left(\rho_t O_I(\theta) \right) = \mathrm{Tr}\left(\rho_\reft O_I(\theta) \right).
  \label{eq: conditions respected}
\end{align}
Thus, we have again obtained the coarse-graining conditions.
\red{As explained before, conversely, maximizing the entropy with those conditions reveals that $\rho_\reft$ must take the form of \eqref{eq: rho_ref}.}
Under these conditions, \eqref{eq: ineq} is reduced to
\begin{align}
  S(\rho_\reft) \geq S(\rho_0).
  \label{eq: second law in CFT}
\end{align}
Thus, our coarse-grained entropy $S(\rho_\reft)$ never gets smaller than the initial value.
Note that the initial state is not arbitrary.

Although not written down, the set of differential equations for $\beta_t$, $\omega^A_t$, and $j^I_t$, i.e, the differential equations to determine $\rho_\reft$, can be derived from the $t$-derivative of \eqref{eq: conditions respected}.
The solution is unique under the initial condition \eqref{eq: initial cond}.
However, we do not argue this point anymore, as the our target is the application to gravity.

\subsubsection{Purification}
For later convenience, we consider purifying $\rho_t$.
To purify $\rho_0$, we bring a copy of the QFT, and name the original one $\mathrm{QFT}_\mathrm{R}$ and the copied one $\mathrm{QFT}_{\mathrm{L}}$.
Let $\ket{\psi_0}$ be the pure state in $\mathrm{QFT}_{\mathrm{L}}\otimes \mathrm{QFT}_{\mathrm{R}}$ given by
\begin{align}
  \ket{\psi_0} = \frac{1}{\sqrt{Z_0}}\sum_n \ket n_\mathrm{L}\otimes e^{-\beta \tilde H_{\mathrm{R}}/2} \ket n_\mathrm{R},
  \label{eq: TFD}
\end{align}
where $\{\ket{n}\}$ is any orthonormal basis, and $\tilde H$ is defined to be,
\begin{align}
  \tilde H := H - \omega^A P_A - \int \d ^{d-1}\theta \sqrt{-\sigma} \, j^I(\theta) O_I(\theta),
  \label{eq: tilde H}
\end{align}
which appeared in the exponent of \eqref{eq: rho_0}.
We see $\rho_0 = \mathrm{Tr}\ketbra{\psi_0}{\psi_0}$, and hence $\ket{\psi_0}$ is the purified state of $\rho_0$.
Here, note that $\tilde H$ is Hermitian due to $\rho_0^\dagger = \rho_0$.

The time evolution of the total system is defined by
\begin{align}
  U_{\mathrm{LR}}(t) = V_{L}(t)\otimes U_R(t),
\end{align}
where $U_{\mathrm{R}}(t)$ is the one in \eqref{eq: rho_t}, and $V_{\mathrm{L}}(t)$ is any unitary operator.
With this, $\ket{\psi_0}$ evolves to
\begin{align}
  \ket{\psi_t} := U_{\mathrm{LR}}(t)\ket{\psi_0}  = \frac{1}{\sqrt{Z_0}} \sum_n V_\mathrm{L}(t) \ket n_\mathrm{L}\otimes U_{\mathrm{R}}(t) e^{-\beta \tilde H_\mathrm{R}/2} \ket {n}_\mathrm{R}.
  \label{eq: evolved TFD}
\end{align}
For any $V_{\mathrm{L}}$ and any operator $O$, we can show
\begin{align}
  &\mathrm{Tr}_{\mathrm{L}} \ketbra{\psi_t}{\psi_t} = \rho_t,\\
  &\braket{\psi_t|O_\mathrm{R}|\psi_t} = \mathrm{Tr}\left(\rho_t O \right).
  \label{eq: left trace out}
\end{align}
Since our focus is $\mathrm{QFT}_{\mathrm{R}}$ and $V_{\mathrm{L}}$ is any, we hereafter set 
\begin{align}
  V(t) = e^{-iHt},\label{eq: V(t) specified}
\end{align}
while continuing using the notation $V$ even below.

\subsection{Path integral representation}
Here, we write down the generating functionals for $\rho_\reft$ and \eqref{eq: left trace out} to use the GKPW formula in section \ref{sec: gravity}.
\subsubsection{Coarse-grained state}
Let us start with
\begin{align}
  \rho[\Gamma] := \frac{1}{Z[\Gamma]} \exp\left[-B \left(H - \Omega^A P_A - \int \d ^{d-1}\theta \, \sqrt{-\sigma} J^I(\theta) O_I(\theta) \right) \right],
   \label{eq: rho_Gamma}
\end{align}
where $\Gamma := (B,\Omega,J)$.
First, notice that
\begin{align}
  H - \Omega^A P_A = \int\d ^{d-1}\theta\sqrt{-\sigma}\, u_a T^a_b \xi^b,\qquad \xi^a := u^a - \Omega^A e^a_A.
\end{align}
From the viewpoint of the ADM formalism, this is the Hamiltonian when the time direction is chosen as $\xi^a$.
In other words, it is the Hamiltonian on the background given by
\begin{align}
  \d s^2 = - \d t^2 + \sigma_{AB}(\d \theta^A - \Omega^A\d t)(\d \theta^B - \Omega^B\d t).
\end{align}
Then, the partition function in \eqref{eq: rho_Gamma} is expressed as
\begin{align}
	Z[\Gamma] = \oint \mathcal{D}\varphi\, \exp \left[-\int_0^B \d\tau \int d^{d-1}\theta \sqrt{\tilde \sigma}\left(\mathcal{L}_{\mathrm{E}} - J^I(\theta)O_I(\tau, \theta) \right)  \right],
	\label{eq: Z[Gamma]}
\end{align}
where $\mathcal{L}_{\mathrm{E}}$ is the Euclid Lagrangian on
\begin{align}
  \d s^2 = \tilde \sigma_{ab}\d z^a \d z^b = \d \tau^2 + \sigma_{AB}(\d \theta^A + i \Omega ^A \d \tau)(\d \theta^B + i \Omega ^B \d \tau).
  \label{eq: Euclid twisted bdy metric}
\end{align}
In the path integral, $\varphi$ means the collection of the elemental fields, and the (anti-)periodic boundary condition is imposed on $\varphi$ as $\tau + B \sim \tau$.
The normalization factor $Z_t$ in \eqref{eq: rho_ref} is equal to $Z[\Gamma_t]$ with $\Gamma_t = (\beta_t,\omega_t,j_t)$.
The expectation values are generated as
\begin{align}
  \mathrm{Tr}\left(\rho[\Gamma] O_I(\theta) \right) = \frac{1}{B\sqrt{\tilde \sigma}} \frac{\delta}{\delta  J^I(\theta)}\ln Z[\Gamma].
  \label{eq: generating Euclid values}
\end{align}

By a coordinate transformation from $\theta^A$ to $\vartheta^A = \theta^A + i \Omega^A \tau$, the metric is changed to
\begin{align}
  \d s^2 = \d \tau^2 + \sigma_{AB}\d \vartheta^A \d \vartheta^B.
  \label{eq: Euclid static bdy metric}
\end{align}
In this coordinate, the periodicity condition for $\tau$ is modified to $(\tau + B ,\vartheta^A + i \Omega^A B) \sim (\tau, \vartheta^A)$, and additionally, the source term could depend on $t$.
In the following, we will rather use the coordinate \eqref{eq: Euclid twisted bdy metric}.
The choice of \eqref{eq: Euclid static bdy metric} is discussed in section \ref{subsubsec: shift vector}.

\begin{figure}
	\centering
	\begin{minipage}{0.25\columnwidth}
		\centering
		\includegraphics[height = 3cm]{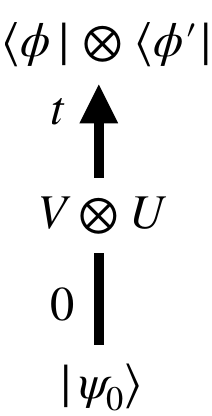}
	\end{minipage}
	\hspace{12pt}
	$=$
	\hspace{24pt}
	\begin{minipage}{0.4\columnwidth}
		\centering
		\includegraphics[height = 6cm]{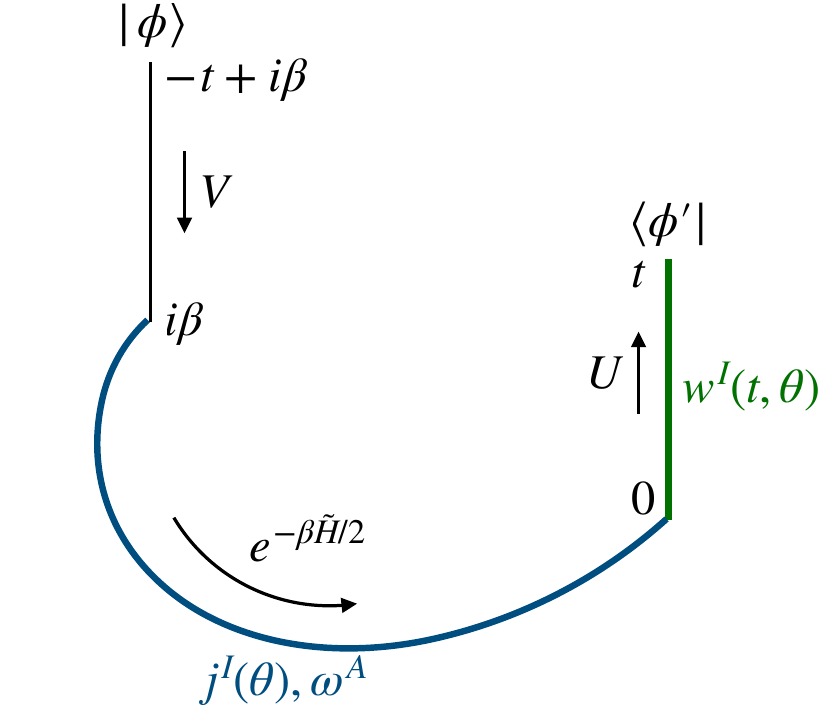}
	\end{minipage}
	\caption{
	(Left) The evolution of \eqref{eq: TFD} to \eqref{eq: evolved TFD}.
	(Right) The contour $C$ in \eqref{eq: complex path int} to give the wave function of \eqref{eq: evolved TFD}.
	The two vertical lines are the Lorentzian time-evolution over $[0,t]$ by $V$ and $U$, while the half curve is the Euclidean evolution over $[0,\beta/2]$.
	In the figure, the Lorentzian time direction of $V$ is drawn opposite to $U$, to match with figure \ref{fig: Lorentzian bulk}.
	}
	\label{fig: contour C}
\end{figure}

\subsubsection{Original state}
Next, we find the path integral representation of the functional to generate \eqref{eq: left trace out}.
As will be shown in appendix \ref{app: derivation}, the following holds for $\ket{\psi_t}$\footnote{
Equation \eqref{eq: wavefunction of TFD} does not hold as it is, but as explained in appendix \ref{app: derivation}, there is no problem in proceeding with \eqref{eq: wavefunction of TFD}.
}:
\begin{align}
  {}_{\mathrm{L}}\hspace{-2.5pt} \bra{\phi}\otimes {}_{\mathrm{R}}\hspace{-2.5pt}\bra{\phi'} V_{\mathrm{L}}(t) \otimes U_{\mathrm{R}}(t) \left(\sum_n \ket n_\mathrm{L}\otimes e^{-\beta \tilde H_\mathrm{R}/2} \ket {n}_\mathrm{R} \right)
  = \braket{\phi'| U(t) e^{-\beta \tilde H/2} V(t) | \phi}.
  \label{eq: wavefunction of TFD}
\end{align}
Note that the l.h.s.\ is proportional to the wave function of the thermofield double state \eqref{eq: evolved TFD}, and that the r.h.s.\ is written in terms of a single QFT.
The r.h.s.\ can be expressed as
\begin{align}
  \braket{\phi'| U(t) e^{-\beta \tilde H/2} V(t) | \phi}
  =
  \int_{\phi}^{\phi'} \mathcal{D}\varphi\, e^{ i I_{\mathrm{QFT}}[\varphi; C]},
  \label{eq: complex path int}
\end{align}
where $C$ is the contour depicted in figure \ref{fig: contour C}; first, $+t$ in the real direction, then $-i\beta$ in the pure imaginary direction, and finally $+t$ again in the real direction.
The metric of the Lorentzian parts is \eqref{eq: boundary metric}, and the one for the Euclidean part is \eqref{eq: Euclid twisted bdy metric} with the replacement $\Omega^A \to \omega^A$.
Regarding the continuity, the induced metric on constant time surface is the same between both metrics, but the extrinsic curvature is not analytically continuous.
This is not a problem however, because there is no dynamical gravity on the QFT.

\begin{figure}
	\centering
	\includegraphics[height = 6cm]{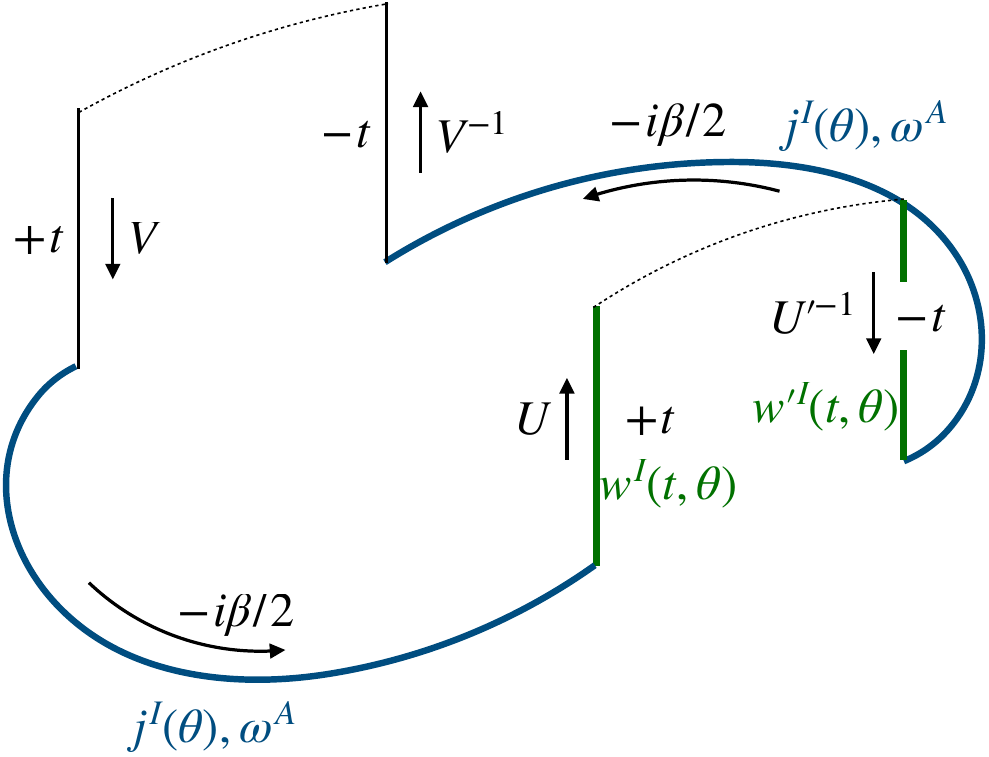}
	\caption{
	The contour $C'$ for the generating functional \eqref{eq: Lorentz generating functional}.
	The source $w'$ in the inverse contour is different from $w$, and $U'$ is the time-evolution operator with source $w'$.
	Note again that the Lorentzian time direction of $V$ and $V^{-1}$ is taken opposite to $U$ and $U'^{-1}$.
	}
	\label{fig: generator contour}
\end{figure}

The generating functional to compute \eqref{eq: left trace out} is the path integral over the closed contour $C'$ in figure \ref{fig: generator contour}:
\begin{align}
  Z_{C'}[w,w'] = \oint \mathcal{D}\varphi\, e^{iI_{\mathrm{QFT}}[\varphi; C']}.
  \label{eq: Lorentz generating functional}
\end{align}
In the arguments, $j$ and $\omega$ have been omitted, as they will be fixed hereafter.
Since $C'$ is closed, the path integral is subject to the (anti-)periodic boundary condition.
This path integral corresponds to 
\begin{align}
  \mathrm{Tr}\left[\left(U'(t) e^{-\beta \tilde H/2} V(t) \right)^\dagger U(t) e^{-\beta \tilde H/2} V(t) \right],
\end{align}
and thus when $w' = w$, it is equal to $ Z_0$, i.e, $Z_{C'}[w,w] = Z_0$.
The relation we need is
\begin{align}
   \mathrm{Tr}\left(\rho_t O_I(\theta) \right) =\l. \frac{1}{i\sqrt{-\sigma}} \frac{\delta}{\delta w^I(t,\theta)}\ln Z_{C'}[w,w']\r|_{w'\to w}.
   \label{eq: functional derivative in CFT}
\end{align}

To compute the expectation values of $H$ and $P_A$, we need to include the stress tensor $T_{ab}(z)$ in the source term of the Lagrangian, coupled to the source $\delta \sigma^{ab}$ as $(1/2)\delta \sigma^{ab} T_{ab}$.
This corresponds to that the Lagrangian \eqref{eq: CFT Lagrangian}, which defines the evolution $U(t)$, is replaced with
\begin{align}
  I_{\mathrm{QFT}}[w, \delta \sigma] 
  &= I_{\mathrm{QFT}}[w] - \frac{1}{2}\int \d ^{d}z\sqrt{-\sigma}\,\delta \sigma^{ab}(z) T_{ab}(z)\nonumber\\
  &= \int\d ^{d}z\, \sqrt{-\mathrm{det}\left(\sigma_{ab} + \delta \sigma_{ab} \right) } \mathcal{L}[\sigma_{ab} + \delta \sigma_{ab}] + \int\d ^{d}z\sqrt{-\sigma}\, w^I(z) O_I(z),
  \label{eq: source to metric}
\end{align}
where $\delta \sigma_{ab} := - \sigma_{ac} \sigma_{bd} \delta \sigma^{cd}$.
Note that the metric in the source term is \textit{not} perturbed.
In the following, we include $-\delta \sigma^{ab} /2$ in $\{w^I\}$, so that we can deal with them collectively.
In the expression \eqref{eq: functional derivative in CFT}, we take $\delta \sigma \to 0$ together with $w' \to w$.

\section{Description in gravity}\label{sec: gravity}
Since the path integral representations \eqref{eq: Z[Gamma]} and \eqref{eq: Lorentz generating functional} have been obtained, the GKPW formula enables us to write them in terms of gravity.
Especially, \eqref{eq: Z[Gamma]} corresponds to a classical Euclidean gravity solution, and we will derive its thermodynamic laws.
Our coarse-grained entropy is, in the classical limit, equal to the horizon area of the Euclidean cigar geometry (figure \ref{fig: cigar}).
The application to non-AdS cases is discussed at the end of this section.

\subsection{Setup}
Here, a $d$-dimensional holographic CFT is considered.
The contents in section \ref{sec: CFT} apply as they are, and we use the holographic dictionary to rewrite them in the gravitational language.
Particularly, we find the dual descriptions of the conditions \eqref{eq: conditions respected} and the coarse-grained entropy $S(\rho_\reft)$.

\begin{figure}
	\centering
	\includegraphics[height = 7cm]{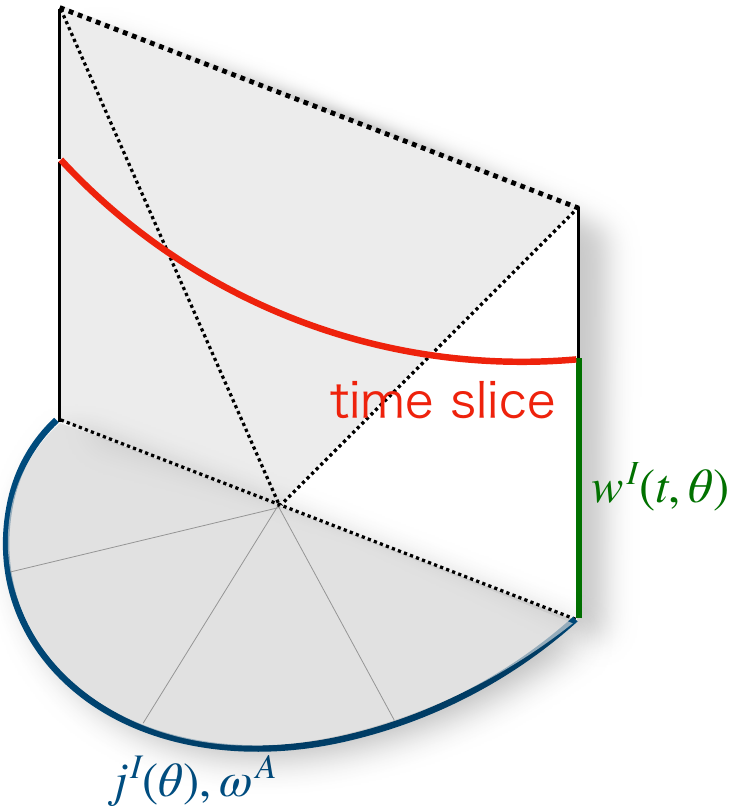}
	\caption{The bulk spacetime. 
	The half disc depicts the Euclidean time evolution over $[0,\beta/2]$ to prepare the initial configuration at $t=0$, which evolves to the Lorentzian part drawn vertically.
	The dotted diagonal lines are the event horizons.
	This classical solution is constructed by properly joining the saddle point of each part \cite{Skenderis:2008dg}.}
	\label{fig: Lorentzian bulk}
\end{figure}

The bulk configuration dual to \eqref{eq: Lorentz generating functional} is constructed by the common procedure \cite{Maldacena:2001kr, Skenderis:2008dg}.
As figure 12 of \cite{Skenderis:2008dg}, the resulting spacetime fills the bulk of figure \ref{fig: generator contour} as shown in figure \ref{fig: Lorentzian bulk}.
(In this figure, only the forward half is drawn, and it is up to the time slice drawn red that corresponds to the forward half of figure \ref{fig: generator contour}.)
Our target is the unshaded single-sided black hole spacetime.

Our setup and assumptions in the dual gravity are as follows:
\vspace{-0.5\baselineskip}
\begin{itemize}[leftmargin=*]
  \setlength{\parskip}{0cm}
  \setlength{\itemsep}{0cm}
  \setlength{\itemindent}{-0pt}
	\item We consider that the bulk $M$ is governed by the Einstein gravity,
			\begin{align}
			  I_{\mathrm{grav}} = \frac{1}{16\pi G}\int_M\d^{d+1}x\sqrt{-g}\, (R - 2\Lambda) + \frac{1}{8\pi G}\int_{\partial M} \d^d z\sqrt{-\gamma}\, K + I_{\mathrm{matter}} + I_{\mathrm{ct}},
			  \label{eq: gravity action}
			\end{align}
			where $I_{\mathrm{matter}}$ is the matter Lagrangian, $I_{\mathrm{ct}}$ is the counterterm both for gravity and matter, and $K_{ab}$ is the extrinsic curvature of the boundary and $K$ is its trace, $K := \gamma^{ab}K_{ab}$. 
			We assume that $I_{\mathrm{matter}}$ does not contain any boundary term.
	\item We write the bulk metric as
			\begin{align}
			  \d s^2 = g_{\mu\nu }\d x^\mu \d x^\nu  = - N^2 \d t^2 + N_r^2 \d r^2 + c_{AB}\left(\d \theta^A + N^A \d t \right)\left(\d \theta^B + N^B \d t \right),
			  \label{eq: bulk Lorentzian metric}
			\end{align}
			with the gauge-fixing condition $g_{rt} = g_{r A} = 0$.
			Throughout this paper, $\mu,\nu,\cdots$ are used for the bulk spacetime coordinates, $A,B,\cdots$ are for $\theta$-coordinates, and $a,b,\cdots$ are for the combination of $(t, \theta^A)$.
	\item With the unit normal $\mathtt{u}_\mu \propto \partial_\mu t$ and the pull-back $e_A^\mu$, the vector of time direction is expressed as
			\begin{align}
			 \mathtt{t}^\mu \partial_\mu :=  \partial_t = (N \mathtt{u}^\mu + N^A e^\mu_A) \partial_\mu = N \mathtt{u}^a \partial_a + N^A \partial_A.
			\end{align}
	\item The boundary $\partial M$ is located at $r\to \infty$, where the induced metric of the Lorentzian part becomes \eqref{eq: boundary metric} after the conformal factor is removed:
			\begin{align}
			  \d s^2 \sim \frac{r^2}{L^2}\left(- \d t^2 + \sigma_{AB}\d \theta^A \d \theta^B \right) + N_r^{2}\d r^2.
			  \label{eq: asymptotic metric}
			\end{align}
			The induced metric on $r = \mathrm{const}$ is $\gamma_{ab}$, which asymptotes to the first term above.
	\item The on-shell value of the action is expressed as $I_{\mathrm{grav}}[w,w']$. 
			The argument $w$ is the one appeared in \eqref{eq: Lorentz generating functional}, and related to the asymptotic value of the bulk fields $\Phi^I$ (including $g_{\mu\nu }$) as
			\begin{align}
  			 \Phi^I (x) \sim  \left(\frac{R}{L} \right)^{\Delta_I - d} w^I(z) =: W^I(z),
			\end{align}
			where $R$ is the IR cut-off to be sent $R\to \infty$, and $\Delta_I$ is the conformal dimension of $O_I$.\footnote{
			In the bulk definition, $\Delta_I$ is determined by the asymptotic $r$-dependence of $\Phi^I$.
			}
			The argument $w'$ is related to $W'$ in the same say in the backward evolution of figure \ref{fig: generator contour}.
	\item The counterterm $I_{\mathrm{ct}}$ is usually determined as \cite{Balasubramanian:1999re, deHaro:2000vlm}.
\end{itemize}
\vspace{-0.5\baselineskip}

\subsection{Coarse-graining conditions in the bulk}
According to the the GKPW formula, \eqref{eq: functional derivative in CFT} is rewritten as
\begin{align}
  \mathrm{Tr}\left(\rho_t O_I(\theta) \right) 
  =
   \lim_{R\to \infty}
  \left(\frac{R}{L} \right)^{\Delta_I} \l. \frac{1}{\sqrt{-\gamma}} \frac{\delta I_{\mathrm{grav}}[w,w']}{\delta W^I(t,\theta)}\r|_{w'\to w}.
  \label{eq: o by I_grav}
\end{align}
For the stress tensor, in particular, we have
\begin{align}
  \mathrm{Tr}\left(\rho_t T_{ab}(\theta) \right) = \lim_{R\to \infty} \left(\frac{R}{L} \right)^{d-2} Y_{ab}(z),
  \label{eq: stress and BY}
\end{align}
where $Y_{ab}$ is the Brown-York tensor defined as \cite{Balasubramanian:1999re}
\begin{align}
  Y_{ab} := -\frac{2}{\sqrt{-\gamma}}\frac{\delta I_{\mathrm{grav}}}{\delta \gamma^{ab}} = \frac{1}{8\pi G}\left(K \gamma_{ab} - K_{ab} \right) - \frac{2}{\sqrt{-\gamma}} \frac{\delta I_{\mathrm{ct}}}{\delta \gamma^{ab}}.
  \label{eq: BY tensor}
\end{align}

Therefore, all we need is
\begin{align}
  \delta_w I_{\mathrm{grav}}[w] :=& I_{\mathrm{grav}}[w + \delta w,w] - I_{\mathrm{grav}}[w,w]\nonumber\\
  =& \int \d ^{d+1}x\, (\mathrm{EOM})_I \delta_w \Phi^I
  + \int \d ^{d}z \sqrt{-\gamma}\, \pi_I(z) \delta W^I(z).
  \label{eq: bulk variation}
\end{align}
Since we are interested in the on-shell variation, only the boundary terms remain.
By definition, the Brown-York tensor is considered as
\begin{align}
  \pi_I = Y_{ab}\qquad
  \mathrm{for}\qquad
  \delta W^I = -\delta \gamma^{ab}/2.
\end{align}
When $I_{\mathrm{matter}}$ takes the form $\int \d ^{d+1} x\sqrt{-g}\, \mathcal{L}_{\mathrm{matter}}(\Phi^{I'}, \nabla_\mu \Phi^{I'};g)$ ($I'$ does not contain $g_{\mu\nu }$), $\pi_{I'}$ for the matter fields are written as
\begin{align}
  \pi_{I'} = r_{\mu} \frac{\partial \mathcal{L}_{\mathrm{matter}}}{\partial \left(\nabla_{\mu} \Phi^{I'} \right)} + \frac{1}{\sqrt{-\gamma}} \frac{\delta I_{\mathrm{ct}}}{\delta W^{I'}} \qquad
  \mathrm{for}\qquad
  \delta W^{I'} = \delta \Phi^{I'}|_{r=R},
  \label{eq: matter pi}
\end{align}
where $r_\mu = N_r^{-1} \delta_\mu ^r$ is the unit normal of $r = \mathrm{const}$.
As seen above, $\pi_I$ is like the ``canonical conjugate momentum" (up to counterterm contributions) if $r$ were viewed as the time coordinate.
This viewpoint agrees with the fact that differentiating the on-shell action with the final position generates the canonical conjugate momentum.

Therefore, \eqref{eq: o by I_grav} is reduced to
\begin{align}
  \mathrm{Tr}\left(\rho_t O_{I'}(\theta) \right) = \lim_{R\to \infty} \left(\frac{R}{L} \right)^{\Delta_{I'}} \pi_{I'}(t,\theta) =:o_{I',t}(\theta),
  \label{eq: rho_t O}
\end{align}
and \eqref{eq: bulk variation} reproduces the on-shell variation formula:
\begin{align}
  \delta_w I_{\mathrm{grav}}[w] = \int \d ^{d}z\sqrt{-\sigma}\, o_{I',t}(\theta)\delta w^{I'}(t,\theta) - \frac{1}{2}\int \d ^{d}z \sqrt{-\gamma}\, Y_{ab}(t,\theta) \delta \gamma^{ab}(t,\theta).
  \label{eq: simple on-shell variation}
\end{align}
For $H$ and $P_A$, the following holds from \eqref{eq: H and P} and \eqref{eq: stress and BY}:
\begin{align}
  &\mathrm{Tr}\left(\rho_t H \right) = \lim_{R\to \infty} \left(\frac{R}{L} \right)^{d-2} \int\d ^{d-1}\theta \sqrt{-\sigma}\, u^a Y_{ab} t^b =  \lim_{R\to \infty} \int\d ^{d-1}\theta \sqrt{c}\, N\mathtt{u}^a Y_{ab} \mathtt{u}^b=:h_t,\label{eq: Lorentz mass} \\
  &\mathrm{Tr}\left(\rho_t P_A \right) = \lim_{R\to \infty} \left(\frac{R}{L} \right)^{d-2} \int\d ^{d-1}\theta \sqrt{-\sigma}\, u^a Y_{ab} e^b_A =  \lim_{R\to \infty} \int\d ^{d-1}\theta \sqrt{c}\, \mathtt{u}^a Y_{ab} e^b_A =: p_{A,t}.\label{eq: Lorentz momentum}
\end{align}
Here, we have used $u^a = t^a \sim N \mathtt{u}^a$ as $r \to \infty$, which follows from \eqref{eq: asymptotic metric}.

It should be noted that $o_{I'}$ and $Y_{ab}$ are evaluated by $I_{\mathrm{\mathrm{grav}}}[w,w]$ ($w'$ is set to be $w$).
According to \cite{Skenderis:2008dg}, the bulk in such a case will be simply constructed as follows.
First, we cut into half the dominant Euclidean black hole solution dual to $Z_0$ (see also next paragraph), which provides the Euclidean ``half disc" in figure \ref{fig: Lorentzian bulk}.
Next, we solve the initial value problem from $t = 0$, with the initial configuration given by analytically continuing the Euclidean solution.
The initial values obtained in this way are compatible with constrains (e.g.\ Hamiltonian constraint).
In solving this problem toward time $t$, the fields are subject to the Neumann boundary conditions specified by $w^I$ on $\partial M$.
The backward evolution part is simply obtained by time-reversing it.

Next, we move on to the dual gravitational description for $\rho_\reft$ in \eqref{eq: rho_ref}.
We assume that the bulk solution that dominates $Z[\Gamma]$ does not depend on the imaginary time (i.e, stationary), and let $I_{\mathrm{grav}}^{(\mathrm{E})}[\Gamma]$ be the bulk Euclidean action evaluated by the dominant saddle point for given $\Gamma$.\footnote{
Although phase transitions can happen depending on $\Gamma$, such as the Hawking-Page transition \cite{cmp/1103922135}, we formally perform the following calculations.
}
Similarly to the Lorentzian case, the same formulae hold for the Euclidean solution:
\begin{align}
   &\mathrm{Tr}\left(\rho[\Gamma] O_{I'}(\theta) \right) = \lim_{R\to \infty}\left(\frac{R}{L} \right)^{\Delta_{I'}} \pi_{I'}(\theta) =: o_{I'}[\Gamma](\theta),\\
  &\mathrm{Tr}\left(\rho[\Gamma] H \right) =  \lim_{R\to \infty} \int\d ^{d-1}\theta \sqrt{\tilde c}\, N\mathtt{u}^a Y_{ab} \mathtt{u}^b =:h[\Gamma],\label{eq: Euclidean Hamiltonian expectation} \\
  &\mathrm{Tr}\left(\rho[\Gamma] P_A \right) =  \lim_{R\to \infty} \int\d ^{d-1}\theta \sqrt{\tilde c}\, \mathtt{u}^a Y_{ab} e^b_A =:p_A[\Gamma],\label{eq: Euclidean momentum expectation}
\end{align}
where, $\pi_{I'}$ here is defined as
\begin{align}
  \pi_{I'}(\theta) := -\frac{1}{B\sqrt{\tilde \gamma}}\frac{\delta I_{\mathrm{grav}}^{\left(\mathrm{E} \right)}[\Gamma]}{\delta \mathcal{J}^I(\theta)},\qquad
  \mathcal{J}^I(\theta) := \left(\frac{R}{L} \right)^{\Delta_{I'}-d}J^I(\theta).
  \label{eq: Euclidean pi}
\end{align}
The tildes on the metrics indicates the Euclidean signature.
As opposed to the Lorentzian case, the time vector is twisted as $\tau^a = u^a + i\Omega^A e_A^a$ due to $\Omega$ in \eqref{eq: Euclid twisted bdy metric}.
However, in order to measure $H$ rather than the Hamiltonian that generates $\tau$-translation of \eqref{eq: Euclid twisted bdy metric}, we had to choose $u^a T_{ab} u^b$, not $u^a T_{ab} \tau^b$.
This point has already been taken into account in \eqref{eq: Euclidean Hamiltonian expectation}.

From the above, \eqref{eq: conditions respected} is equivalent in the bulk to finding the set $\Gamma = (B,\Omega,J)$ such that 
\begin{align}
  o_{I'}[\Gamma] = o_{I',t}\qquad
  h[\Gamma] = h_t,\qquad
  p_{A}[\Gamma] = p_{A,t}.
  \label{eq: conditions respected in bulk}
\end{align}
We write the solution as $\Gamma_t = (\beta_t, \omega_t, j_t)$.
The first condition is equivalent to equating the leading behaviors of $\pi_I$.

\subsection{Coarse-grained entropy}
\begin{figure}
	\centering
	\includegraphics[height = 4cm]{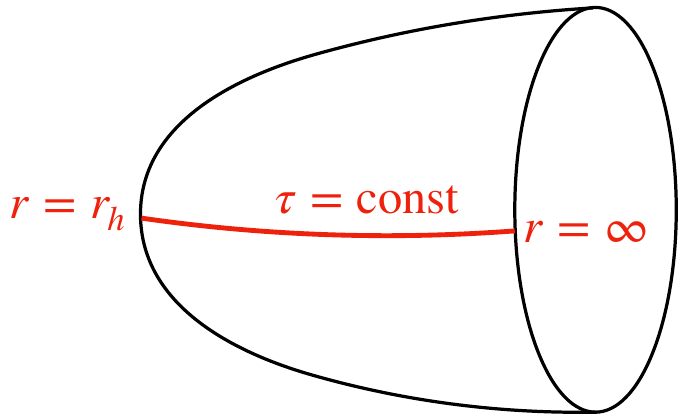}
	\caption{The Euclidean cigar geometry and a time slice $\tau = \mathrm{const}$.}
	\label{fig: cigar}
\end{figure}

After $\Gamma_t$ is found for each time $t$, our coarse-grained entropy can be computed in the gravity.
We write the bulk metric as
\begin{align}
  \d s^2 = \tilde g_{\mu\nu }\d x^\mu \d x^\nu  = \tilde N^2 \d \tau^2 + \tilde N_r^2 \d r^2 + \tilde c_{AB}\left(\d \theta^A + i \tilde N^A \d \tau \right)\left(\d \theta^B +i \tilde N^B \d \tau \right),
  \label{eq: bulk Euclidean metric}
\end{align}
similar to \eqref{eq: bulk Lorentzian metric}.
We write the metric on $r = \mathrm{const}$ as $\tilde \gamma_{ab}$.
This metric asymptotes to \eqref{eq: Euclid twisted bdy metric} as
\begin{align}
  \tilde N \sim \frac{r}{L},\qquad \tilde N^A \sim \Omega^A,\qquad \tilde c_{AB} \sim \frac{r^2}{L^2} \sigma_{AB},\qquad (r \to \infty).
  \label{eq: Euclid asymptotic metric}
\end{align}
We suppose $\tilde N_r|_{r= r_h} = 0$ and $\partial_r \tilde N_r|_{r= r_h} \neq 0$ for some $r_h> 0$, as figure \ref{fig: cigar}.\footnote{
If $\tilde N \neq 0$ anywhere, the entropy just vanishes and such a case is not our interest.
}
In this case by a coordinate transformation $\d \rho = N_r \d r$ ($\rho(r_h) = 0$), the metric near $r = r_h$ must behave as
\begin{align}
  \d s^2 \sim \d \rho^2 + \left(\frac{2\pi}{B} \rho \right)^2 \d \tau^2 + \tilde c_{AB}\d \theta^A \d \theta^B,
  \label{eq: static near horizon metric}
\end{align}
so as to avoid the conical singularity.
Note that the shift vectors must also vanish (see below \eqref{eq: Euclid twisted bdy metric} and also section \ref{subsubsec: shift vector}).

First, the entropy of $\rho[\Gamma]$ in \eqref{eq: rho_Gamma} is
\begin{align}
  S(\rho[\Gamma]) &= -\mathrm{Tr}\left(\rho[\Gamma]\ln \rho[\Gamma] \right)\nonumber\\
   &= \ln Z[\Gamma] + B\, \mathrm{Tr}\left(\rho[\Gamma] \left\{H - \Omega^A P_A -\int \d ^{d-1}\theta \, \sqrt{-\sigma} J^I(\theta) O_I(\theta)  \right\} \right)\label{eq: entropy with expectation values}\\
   &= B^2\frac{\partial}{\partial B} \left(-B^{-1}\ln Z[\Gamma] \right)\label{eq: entropy with B-derivative}.
\end{align}
Using the holographic dictionary, $-\ln Z[\Gamma] = I_{\mathrm{grav}}^{(\mathrm{E})}[\Gamma]$, we obtain
\begin{align}
  S(\rho[\Gamma]) = I_{\mathrm{grav}}^{(\mathrm{E})}[\Gamma] - B\frac{\partial I_{\mathrm{grav}}^{(\mathrm{E})}[\Gamma]}{\partial B} = B \hat \d_B I_{\mathrm{grav}}^{(\mathrm{E})}[\Gamma],
  \label{eq: entropy by d_B}
\end{align}
where $\hat \d_B$ is $B$-derivative that acts only on the fields (with the integration range of $\tau$ fixed) and the last equation follows from the $\tau$-independence of the on-shell fields.
Thus, $\hat \d_B I_{\mathrm{grav}}^{(\mathrm{E})}[\Gamma]$ is just the field variation triggered by $B$, and the Euclidean version of \eqref{eq: bulk variation} can be used with $\delta_w \to \hat \d_B$.
Since the bulk fields are subject to boundary conditions $J^I(\theta)$ independent of $B$, \eqref{eq: bulk variation} seems to vanish.
But, the variation perturbs $B$ in \eqref{eq: static near horizon metric}, generating a non-trivial conical singularity.
Due to this, we have \cite{Fursaev:1995ef}
\begin{align}
	-\frac{1}{16\pi G}\int \d ^{d+1}x\, \hat \d_B(\sqrt{\tilde g}\, R) = \frac{1}{4GB}\int_{r_h}\d ^{d-1}\theta\sqrt{\tilde c} + \left(\mbox{regular terms} \right),
\end{align}
by which we obtain
\begin{align}
  S\left(\rho[\Gamma] \right) = \frac{1}{4G}\int_{r_h}\d ^{d-1}\theta\sqrt{\tilde c}.
  \label{eq: area entropy}
\end{align}
The fact that the entropy is given by the area of the tip agrees with \cite{Lewkowycz:2013nqa}, where the replica trick is used.

\red{If necessary, quantum corrections can be computed as in \cite{Barrella:2013wja, Faulkner:2013ana, Engelhardt:2014gca, Almheiri:2019qdq, Solodukhin:1994yz}.
Also, since \eqref{eq: area entropy} is derived from the gravitational path integral $Z[\Gamma]$, the non-perturbative effects such as phase transitions are automatically taken into account when the saddle point approximation is taken for $Z[\Gamma]$.
If the dominant saddle point changes, the entropy is then given by the corresponding Euclidean solution.
In any case, the second law introduced below in \eqref{eq: the second law} must be satisfied.
}

\subsection{Thermodynamic laws}
The second law is guaranteed via the AdS/CFT correspondence by \eqref{eq: second law in CFT}:
\begin{align}
  S(\rho[\Gamma_t]) \geq S(\rho[\Gamma_0]).\label{eq: the second law}
\end{align}
Here, $S(\rho[\Gamma_0])$ coincides with the horizon area of the initial black hole, as we have set $\rho_0 = \rho_{\mathrm{ref},0}$.
Therefore, the horizon area of the auxiliary Euclidean black hole never gets smaller than the initial value.
\red{This second law is a requirement from the AdS/CFT and is expected to provide a thermodynamic constraint on quantum gravity.
As we will see in section 4, it seems to be related to some energy condition.}

Next, let us derive the first law.
To find it, we perform a scaling transformation $\tau_s =  B^{-1}\tau$.
Values in the scaled coordinate is put subscript $s$ below.
The periodicity is modified to $\tau_s + 1\sim \tau_s$, and \eqref{eq: Euclid asymptotic metric} and \eqref{eq: static near horizon metric} are changed to
\begin{align}
  &\tilde N_s \sim B \frac{r}{L},\qquad \tilde N^A_s \sim B\Omega^A,\qquad (r \to \infty),\\
  &\d s^2 \sim \d \rho^2 +  (2\pi \rho)^2 \d  \tau_s^2 + \tilde c_{AB}\d \theta^A \d \theta^B\qquad (r = r_h).
  \label{eq: scaled horizon metric}
\end{align}
In this coordinate, the derivative $\hat \d_B$ does not give any conical singularity, but this time it act on the boundary metric.
In addition, the volume factors transformed as $\sqrt{\tilde \gamma_s} = B\sqrt{\tilde \gamma}$ and $\sqrt{\tilde \sigma_s} = B\sqrt{\tilde \sigma}$ while $Y_{ab}u^au^b$ and $Y_{ab}u^a e^b_A$ are invariant.
Noting those points, we take a variation $\delta \Gamma$ for $I_{\mathrm{grav}}^{(\mathrm{E})}[\Gamma]$ (see \eqref{eq: generating Euclid values}, \eqref{eq: simple on-shell variation} and \eqref{eq: Euclidean pi}):
\begin{align}
  \delta_\Gamma I_{\mathrm{grav}}^{(\mathrm{E})}[\Gamma] =& -\int \d ^{d-1}\theta \sqrt{\tilde \sigma_s}\, o_{I'}[\Gamma](\theta)\delta J^{I'}(\theta) - \frac{1}{2}\int \d ^{d-1}\theta \sqrt{\tilde \gamma_s}\, Y_{ab}(\theta) \delta_\Gamma \gamma^{ab}(\theta),
  \label{eq: scaled variation}\\
 \delta_\Gamma \gamma_{ab} =& \frac{2}{B}\left(\delta B\mathtt{u}_a \mathtt{u}_b - \delta (B\Omega^A)\mathtt{u}_a e_{Ab} \right).\label{eq: delta Gamma to gamma}
\end{align}
Therefore, going back to the original coordinate, we obtain
\begin{align}
  \delta_\Gamma I_{\mathrm{grav}}^{(\mathrm{E})}[\Gamma] =  h[\Gamma] \delta B - p_A[\Gamma] \delta(B\Omega^A) - \int \d ^{d-1}\theta \sqrt{\tilde \sigma}\, \hat o_{I'}[\Gamma](\theta)\delta J^{I'}(\theta),
  \label{eq: unscaled variation}
\end{align}
where we have defined
\begin{align}
  \hat o_{I'}[\Gamma] := B o_{I'}[\Gamma].
\end{align}
In \eqref{eq: unscaled variation}, we have supposed that $\{\Phi^{I'}\}$ includes only fields that are invariant under the scaling.
Soon later, we will see what happens if there are scaling fields such as the Maxwell field.

The entropy function is obtained by the Legendre transformation from $\Gamma$ to $\Xi := (h,p,\hat o)$:
\begin{align}
  S[\Xi] :=& B[\Xi] \left (h - p_A \Omega^A[\Xi] \right ) - \int \d ^{d-1}\theta \sqrt{\tilde \sigma}\, \hat o_{I'}(\theta) J^{I'}[\Xi](\theta) - I_{\mathrm{grav}}^{(\mathrm{E})}[\Gamma[\Xi]]\label{eq: entropy function} \\
  \delta_{\Xi}S[\Xi] =& B[\Xi]\left(\delta h - \Omega^A[\Xi]\delta p_A  \right) - \int \d ^{d-1}\theta \sqrt{\tilde \sigma}\, \delta \hat o_{I'}(\theta) J^{I'}[\Xi](\theta)\label{eq: pre-first law}.
\end{align}
Now, the entropy has a new expression \eqref{eq: entropy function}, which is seemingly different from \eqref{eq: area entropy}.
But, they must be equivalent:
\begin{align}
  S(\rho[\Gamma_t]) = S[\Xi_t],\qquad
  \Xi_t := (h_t, p_t, \hat o_t).
\end{align}
As a matter of fact, \eqref{eq: entropy function} is exactly the same as the corresponding CFT expression \eqref{eq: entropy with expectation values}.
By construction of $\Gamma_t$, the argument $\Xi_t$ is the one given in \eqref{eq: rho_t O}, \eqref{eq: Lorentz mass}, and \eqref{eq: Lorentz momentum} up to the hat symbol.
The first law follows from \eqref{eq: pre-first law}:
\begin{align}
  \dot S[\Xi_t] = B[\Xi_t]\left(\dot h_t - \Omega^A[\Xi_t] \dot p_{A,t}  \right) - \int \d ^{d-1}\theta \sqrt{\tilde \sigma}\, \dot{\hat{o}}_{I',t}(\theta) J^{I'}[\Xi](\theta).
  \label{eq: first law}
\end{align}
This equation contains the local terms from matter fields, in addition to the usual first law.

Finally, let us see how \eqref{eq: unscaled variation} and below are modified when there are scaling fields.
To be concrete, we demonstrate with the time component of $\mathrm{U}(1)$ gauge field $A_0$, whose $J^I$ and $o_I$ we write as $a$ and $q$, respectively.
In this case, \eqref{eq: unscaled variation} is modified to
\begin{align}\label{eq: modified unscaled version}
  \delta_\Gamma I_{\mathrm{grav}}^{(\mathrm{E})}[\Gamma] 
  = \eqref{eq: unscaled variation} - \int \d ^{d-1}\theta \sqrt{\tilde \sigma}\, q[\Gamma](\theta)\delta (Ba(\theta)).
\end{align}
Accordingly, $q$ is added to $\Xi$, and \eqref{eq: entropy function} and \eqref{eq: first law} acquire additional terms:
\begin{align}
  S[\Xi] :=& ~\eqref{eq: entropy function} - B[\Xi] \int \d ^{d-1}\theta \sqrt{\tilde \sigma}\, q(\theta) a[\Xi](\theta),\\
  \dot S[\Xi_t] =& ~\eqref{eq: first law} - B[\Xi_t]\int \d ^{d-1}\theta \sqrt{\tilde \sigma}\, \dot q_t(\theta) a[\Xi_t](\theta).\label{eq: modified first law}
\end{align}
The additional term is the usual term consisting of the $\mathrm{U}(1)$ electric charge and the gauge potential, except that they are locally treated in the current case.
If there are more fields scaled by $\tau_s = B^{-1}\tau$, similar terms are added depending on their transformation rules.

\subsection{Comments on generalizations}
\subsubsection{Non-AdS cases}\label{subsubsec: non-AdS}
We have obtained the definition of our coarse-grained entropy purely in the gravitational language.
If gravity is in general holographic in the sense that the quantum theory behind it is defined by specifying the boundary as assumed in \cite{Lewkowycz:2013nqa}, then we would reach the same results in the classical limit from the saddle point approximation applied to the fomal  path integral of the quantum gravity.
Thus, it is expected that the first and second law also holds for non-AdS cases.

Let us consider the Einstein gravity \eqref{eq: gravity action} on a Lorentzian manifold $M$ with a timelike boundary $\partial M$.
The induced metric on $\partial M$ is assumed static, but the boundary conditions for the other bulk fields can depend on the time.
Then, \eqref{eq: conditions respected in bulk} determines the reference Euclidean solution on each time, whose horizon area is the entropy we want.
In solving \eqref{eq: conditions respected in bulk}, the Brown-York tensor is defined in the same way, and the matching condition of $o_I$ is replaced with the matching of the leading term of $\pi_I$.

\subsubsection{Choice of shift vectors}\label{subsubsec: shift vector}
So far, we have adopted \eqref{eq: Euclid twisted bdy metric} as our gauge choice.
As has been explained, it is possible to choose \eqref{eq: Euclid static bdy metric}, where $\vartheta$ is shifted in the pure imaginary direction along with the $\tau$-periodicity.
If we had taken this gauge, \eqref{eq: static near horizon metric} would be changed to
\begin{align}
  \d s^2 \sim \d \rho^2 + \left(\frac{2\pi}{B} \rho \right)^2 \d \tau^2 + \tilde c_{AB}(\d \vartheta^A - i\Omega^A \d \tau) (\d \vartheta^B - i\Omega^B \d \tau),
\end{align}
from the requirement of removing the conical singularity.
Seemingly different, but physically same results must be reproduced also in this gauge.
A difference, for example, is that the (angular) momenta are measured not on the boundary, but at $r=r_h$.
In fact, $\Omega$ does not appear in \eqref{eq: delta Gamma to gamma}, but in \eqref{eq: static near horizon metric} as above, and hence $B$ enters to \eqref{eq: scaled horizon metric} accompanying $\Omega$ when $\tau$ is scaled.

It was addressed in \cite{Awad:2007me} for Kerr-AdS$_5$ spacetime, how the choice of the shift vectors affect on the first law.
Since we allow matter fields to exist, the story seems to be not so simple as \cite{Awad:2007me}.
In vacuum solutions, for example, the angular momentum (Komar integral) measured at infinity is equal to the one at the horizon by the Stokes' theorem, but different by the volume integral when matter exists.
Nevertheless, the first law will be properly derived in a similar way.

\section{Demonstration in Vaidya-type spacetimes}\label{sec: demo}
In this section, we check thermodynamic laws derived in the previous section for null-ray collapse models.
Since the second law was predicted from the CFT, it provides an energy constraint for those examples.

\subsection{Schwarzschild-AdS\texorpdfstring{$_{d+1}$}{TEXT}}

\begin{figure}
	\centering
	\includegraphics[height = 4cm]{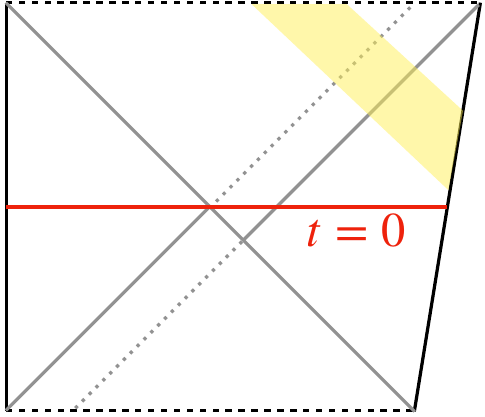}
	\caption{Vaidya-Schwarzschild-AdS$_{d+1}$ spacetime.
	At $t=0$, the Schwarzschild-AdS spacetime is prepared, and null rays exist in the yellow region.}
	\label{fig: Vaidya-AdS}
\end{figure}
Let us start with the simplest case where a null-ray is shot from the boundary to the Schwarzschild-AdS$_{d+1}$ spacetime.
The initial metric is given as
\begin{align}
  \d s^2 = -f(r) \d t^2 + \frac{\d r^2}{f(r)} + r^2 \d\Theta^2,\qquad f(r) := 1 + \frac{r^2}{L^2} - \frac{2 \mu(0)}{r^{d-2}},
\end{align}
where $\d \Theta^2$ is the metric of the unit $(d-1)$-dimensional sphere.
The boundary metric with the conformal factor removed is
\begin{align}
  \d s^2|_{\mathrm{bdy}} = \d t^2 + r^2 \d \Theta^2.\label{eq: bdy metric in AdS}
\end{align}
The metric describing the null ray is
\begin{align}
  \d s^2 = -f(v,r)\d v^2 + 2\d v\d r + r^2 \d \Theta^2,\qquad
  f(v,r) = 1 + \frac{r^2}{L^2} - \frac{2 \mu(v)}{r^{d-2}}.
\end{align}
At the boundary, both time coordinates coincide, namely, $t = v$.
In the bulk, the time coordinate $t$ is properly extended towards the future, but we do not concretely specify the foliation, since only the boundary observables at time $t$ are required in coarse-graining.
Hence, we use $v$ for the time coordinate hereafter.

We take the mass $h_v$ as the only respected observable on the boundary.\footnote{
In principle, we can add boundary values of the matter fields making up the null ray, but it is in general difficult to explicitly construct null ray with some field \cite{Faraoni:2020lte}.
}
The mass $h_v$ depends on the counterterm \cite{Balasubramanian:1999re, deHaro:2000vlm}, but most of time \cite{Balasubramanian:1999re}, it is computed as
\begin{align}
  h_v = \frac{d-1}{8\pi G}\mathrm{Vol}\left(\mathbb{S}^{d-1} \right)\mu (v) + \left(\mbox{independent of $v$} \right).
\end{align}
The Euclidean black hole that minimizes the Euclidean action, having the same mass, will be
\begin{align}
  \d s^2 = f(v,r) \d \tau^2 + \frac{\d r^2}{f(v,r)} + r^2 \d\Theta^2.
\end{align}
Note that $\tau$ is not related to the Lorentzian time $v$, but just the imaginary time for the reference Euclidean solution that we refer to at time $v$.
It is explicitly confirmed that this geometry has the mass $h_v$.
The Lagrange multiplier $\beta_v$ is found to be
\begin{align}
  \beta_v = \frac{4\pi}{\partial_r f(t,r_h(v))},
\end{align}
where $r = r_h(v)$ is the largest root of $f(v, r)$, related to $\mu(v)$ as
\begin{align}
  \mu(v) = \frac{1}{2}r_h(v)^{d-2}\left(1 + \frac{r_h(v)^2}{L^2} \right).
\end{align}

The coarse-grained entropy is the area of the cigar tip,
\begin{align}
  S_v = \frac{\mathrm{Vol}\left(\mathbb{S}^{d-1} \right)}{4 G}r_h(v)^{d-1}.
\end{align}
We see that the first law is satisfied:
\begin{align}
  \dot S_v = \beta_v \dot h_v.
\end{align}
On the other hand, the second law $S_v\geq S_0$ is equivalent to $r_h(v) \geq r_h(0)$, which is not satisfied for generic $\mu(v)$.
We rather see this fact as a constraint from quantum gravity, since this must be satisfied in the dual description.

\begin{figure}
	\centering
	\includegraphics[height = 4cm]{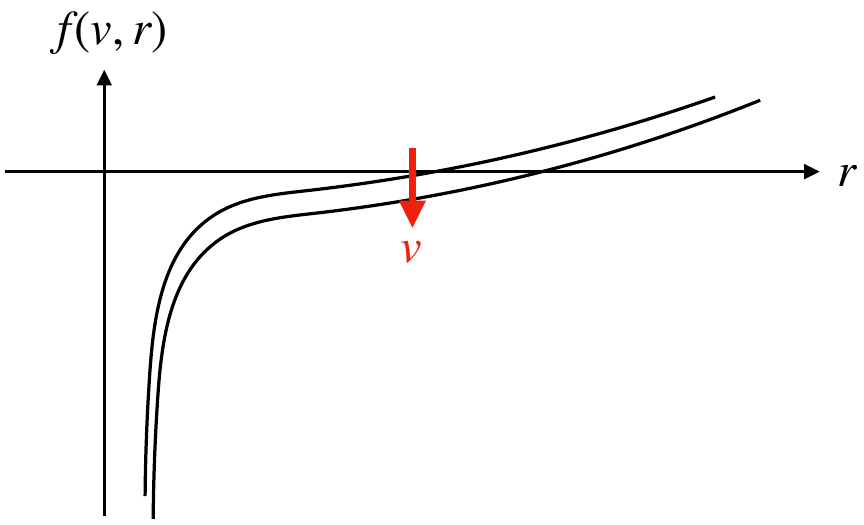}
	\caption{The behavior of $f(v,r)$ under the null energy condition.}
	\label{fig: shift behavior}
\end{figure}

Actually, the second law is satisfied when the null energy condition holds.
The energy flow along a null vector $\ell = \partial_v + (f(v,r)/2)\partial_r$ is
\begin{align}
 T_{\ell \ell} := T_{\mu\nu }\ell^\mu \ell^\nu \propto - \partial_v f(v,r).
\end{align}
If $T_{\ell\ell}\geq 0$, then $f(v,r)$ decreases in $v$, which implies that $r_h(v)$ increases (see figure \ref{fig: shift behavior}).
However, this monotonicity is stronger than required by our second law, and the following \textit{integrated} version is sufficient for only saying $r_h(v) \geq r_h(0)$:
\begin{align}
  \forall v>0,\qquad \int_{0}^v\d v\,T_{\ell \ell} \geq 0.\label{eq: integrated}
\end{align}

\subsection{Rotating BTZ}
Next, we consider the uncharged BTZ spacetime and add the angular momentum $p_\phi$ to the respected observable set.
The metric with null ray is given as
\begin{align}
  \d s^2 &= -N(v,r)^2 \d v^2 + 2 \d v\d r + r^2(\d \phi + N^\phi(v,r)\d v)^2,\\
  N(v,r)^2 &= - 8Gm(v) + \frac{r^2}{L^2} + \frac{16 G^2 j(v)^2}{r^2},\qquad
  N^\phi(v,r) = \frac{4Gj(v)}{r^2}.
\end{align}
The boundary metric is again \eqref{eq: bdy metric in AdS} with $\d \Theta^2 = \d \phi^2$.
We assume that $N(v,r)^2$ has two positive roots, i.e, $L^2m(v)^2 > j(v)^2$, and the larger one is
\begin{align}
  r_h(v) = \left[4L^2 Gm(v)\left(1 + \sqrt{1 - \frac{j(v)^2}{L^2m(v)^2}} \right) \right]^{1/2}.
\end{align}

The counterterm for gravity in $d=2$ \cite{Balasubramanian:1999re, deHaro:2000vlm} is known as
\begin{align}
  I_{\mathrm{ct}} = -\frac{1}{8\pi GL} \int \d^2 z\sqrt{-\gamma},
\end{align}
with which the mass and angular momentum are calculated as
\begin{align}
  h_v = m(v),\qquad
  p_{\phi,v} = j(v).
\end{align}

In considering the reference Euclidean geometry at $v$, we have two parameters $\beta_v$ and $\omega^\phi_v$.
The metric that has $h_v$ and $p_{\phi,v}$, and is compatible with the boundary metric \eqref{eq: Euclid twisted bdy metric}, is found to be
\begin{align}
  \d s^2 =& N(v,r)^2 \d \tau^2 + \frac{\d r^2}{N(v,r)^2} + r^2\left(\d \phi -i \hat N^\phi(v,r) \d \tau \right)^2,\\
  \hat N^\phi(v,r) :=& N^\phi(v,r) - N^\phi(v,r_h(v)).
\end{align}
Thus, the parameters are determined as
\begin{align}
  \beta_v = \frac{4\pi}{\partial_r N(v,r_h(v))^2},\qquad
  \omega^{\phi}_{v} = N^{\phi}(v,r_h(v)).
\end{align}

The coarse-grained entropy is
\begin{align}
  S_v = \frac{\pi}{2 G}r_h(v),
\end{align}
and the first law holds as
\begin{align}
  \dot S_v = \beta_v(\dot h_v - \omega^\phi_v \dot p_{\phi,v}).
\end{align}
The second law means $r_h(v) \geq r_h(0)$.
Again, this is guaranteed under the null energy condition \eqref{eq: integrated} for a null vector $\ell = \partial_v + (N^2/2)\partial_r - N^\phi\partial_\phi$, by the same logic.

\subsection{Reissner-Nordstr\" om (4d flat)}\label{subsec: RN}

\begin{figure}
	\centering
	\includegraphics[height = 4cm]{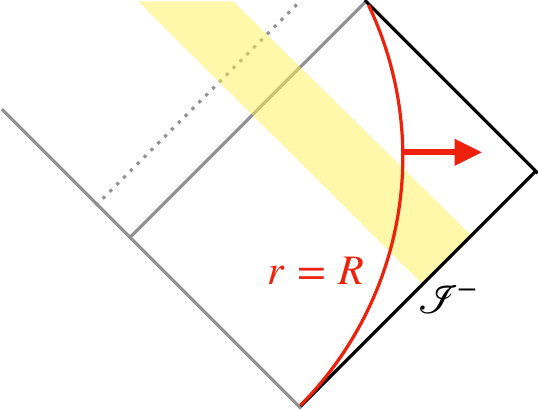}
	\caption{Vaidya-Bonner spacetime.
	The boundary is sent to infinity.}
\end{figure}

The final example is a 4-dimensional charged black hole.
The solution we consider is the Vaidya-Bonner spacetime,
\begin{align}
  \d s^2 =& -N(v,r)^2\d v^2 + 2 \d v \d r + r^2 \d \Theta^2,\qquad
  N(v,r)^2 = 1 - \frac{2G m(v)}{r} + \frac{G q(v)^2}{4 \pi r^2},\\
  A =& -\frac{q(v)}{2\pi r} \left(\d v - \frac{\d r}{N(v,r)^2} \right),
\end{align}
where $\d \Theta^2 = \d \theta^2 + \sin^2\theta \d \phi^2$, and $r_h(v)$ is the largest root of $N(v,r)^2$ given as
\begin{align}
  r_h(v) = Gm(v) + \sqrt{(Gm(v))^2 - \frac{G q(v)^2}{4\pi}},
\end{align}
with the quantity in the square root assumed positive.
We choose $r = R$ as the boundary, whose induced metric is
\begin{align}
  \d s^2|_{\mathrm{bdy}} = - N(v,R)^2 \d v^2 + R^2 \d \Theta^2.
\end{align}
Since this induced metric is time-dependent, we send $R$ to infinity, in order for it to be time-independent.
This infinity corresponds to the past null infinity $\mathscr I^-$.
This time, the respected quantities are the mass and the electric charge.\footnote{
We can also consider \eqref{eq: matter pi} for all components of $A_\mu$, which will make the coarse-graining conditions stronger.
However, as the spacetime is spherically symmetric now, the entropy and its thermodynamic laws will be the same as what we are going to derive below.
}
Observed at $\mathscr I^-$, the mass and charge are
\begin{align}
  h_v = m(v),\qquad
  q_v = q(v),
\end{align}
where the counterterm is, as usual, taken to be the contribution from the Minkowski spacetime.
The mass is the Bondi-Sachs type mass.

The reference Euclidean spacetime is
\begin{align}
  \d s^2 = N(v,r)^2\d \tau^2 + \frac{\d r^2}{N(v,r)^2} + r^2 \d\Theta^2,\qquad
  A = -\frac{q(v)}{2\pi}\left(\frac{1}{r} - \frac{1}{r_h(v)} \right)\d \tau,
\end{align}
with the Lagrange multipliers given by
\begin{align}
  \beta_t = \frac{4\pi}{\partial_r N(v,r_h(v))^2},\qquad
  \lambda_t = \frac{1}{2\pi}A_\tau|_{r\to \infty},
\end{align}
where $\lambda_t$ is the conjugate of $q_v$.\footnote{
Here we take the gauge $A_\tau|_{r=r_h} = 0$ so that $\lambda_t$ becomes a pure boundary value.
Without this gauge, \eqref{eq: modified first law} does not hold as it is, but the term related to $A_\tau|_{r=r_h}$ will appear.
}

The entropy is, hence,
\begin{align}
  S_v = \frac{\pi}{G}r_h(v)^2.
\end{align}
The first law is satisfied in the form \eqref{eq: modified first law}:
\begin{align}
  \dot S_v = \beta_v (\dot h_v -  \lambda_v \dot q_v).
\end{align}
The second law again implies \eqref{eq: integrated}, with $\ell = \partial_v + (N^2/2)\partial_r$.

If we decide not to respect the charge, the resulting entropy must become larger.
This choice is possible when $q(0) = 0$, necessary for $S_0$ to coincide with the initial entropy.
In this case, the reference Euclidean geometry would be
\begin{align}
  \d s^2 = f(v,r )\d \tau^2 + \frac{\d r^2}{f(v,r)} + r^2\d\Theta^2,\qquad
  f(v,r) = 1 - \frac{2 G m(v)}{r}.
\end{align}
Then, the horizon radius is $2Gm(v)$, which is larger than $r_h(v)$, and hence the entropy becomes larger as well.
This reflects the fact that coarse-grained entropy increases when fewer constraints are imposed.

\section{Discussions}
We have introduced a coarse-grained entropy that respects asymptotic values on each time slice, as a new measure of black hole thermodynamics.
The entropy is the horizon area of a specific reference Euclidean black hole, faithful to the first and second laws.
The second law, proven from the CFT, is expected to be a reflection of the thermal (or statistical) nature of quantum gravity.
As a matter of fact, the second law is not always satisfied, providing a constraint on gravity.
Through several examples, we have discovered that the constraint seems to be related to the integrated null energy condition.

The coarse-grained entropy introduced in this paper has its definition both on the boundary and in the bulk.
As a newly established dictionary, this will enable us to study the non-equilibrium process from CFT to gravity, and vice versa.
Although we have focused on the thermodynamics of gravity in this paper, it is of course possible to survey the evolution of the coarse-grained entropy in the CFT from the bulk analysis.

While the first law was derived within the Einstein gravity, we asked for the help of the holographic dictionary in showing the second law.
It is valuable to further survey what the second law means in more complex situations, since the second law seems to imply something that the Einstein theory itself does not tell.
Particularly, we did not specify the origin of the null rays in Vaidya models, but if it were possible, we could add the matter fields to the set of respected quantities.
For this purpose, the solution in \cite{Aniceto:2017gtx}, which offers a null ray collapse model within given fields, will be helpful.

\red{
Also, studying the relation between the coarse-grained entropy and the area of the apparent horizon will be helpful to understand the origin of the second law in the bulk. 
In the Schwarzschild-AdS case in section \ref{sec: demo}, for example, our coarse-grained entropy coincides with the apparent horizon area when the time foliation is specified by the coordinate $v$.
However, the definition of our coarse-grained entropy is essentially different from the apparent horizon; while the apparent horizon is foliation-dependent, our entropy depends only on the boundary time foliation.
To connect them, a proper rule to extend the boundary foliation to the bulk must be established.
A definition of the entropy for dynamical black holes was recently proposed also in \cite{Hollands:2024vbe}, where the apparent horizon appears as well.
One possible future direction is to find a connection between those notions.
}

It is also interesting to consider other ways of coarse-graining to provide some thermodynamic constraints.
In our coarse-graining process, especially, only the boundary observables on the time slice are referred to at each time $t$, and hence it does not tell any difference that happens deep inside the bulk.
To be more finely grained, the coarse-grained state must also respect observables off the time slice.
By doing so, the information of some codimension-$0$ domain on the boundary is kept, preserving some bulk region causally related to it.
This idea has been addressed in the literature \cite{Kelly:2013aja, Engelhardt:2017aux, Engelhardt:2018kcs}.

Finally, one of the aims of non-equilibrium thermodynamics is to describe the relaxation process.
If there is non-equilibrium thermodynamics in gravity, does it also have knowledge about the relaxation of gravitational systems?
In some sense, we can say that the no-hair theorem teaches what states are possible as the final states, after matter has fallen behind the horizon.
However, it is still a mystery how the system evolves to the final state, or whether the fate of the system must be thermodynamically controlled.
We expect that studying non-equilibrium black hole thermodynamics will reveal those problems.

\subsection{Erratum added}
After the publication, I found some ambiguity regarding the dictionary \eqref{eq: source to metric} and \eqref{eq: stress and BY}, which leads some inconsistencies as explained below.\footnote{I thank Sotaro Sugishita for discussion.}

From \eqref{eq: unscaled variation}, we obtain
\begin{align}\label{eq: I variation}
  \delta_B I_{\mathrm{grav}}^{(E)}[\Gamma] = \left(h[\Gamma] - \Omega^A p_A[\Gamma] \right) \delta B,
\end{align}
where $\Gamma$ was defined to be the collection of the Lagrange multipliers.
On the other hand, from the holographic dictionary, $I_{\mathrm{grav}}^{(E)}[\Gamma] = - \ln Z[\Gamma]$, with
\begin{align}
  Z[\Gamma] = \mathrm{Tr}\exp\left[-B\left(H - \Omega^A P_A - \int \d^{d-1}\theta\sqrt{-\sigma}J^I(\theta)O_I(\theta) \right) \right],
\end{align}
we also have
\begin{align}\label{eq: Z variation}
  \delta_B \left(-\ln Z[\Gamma] \right) = \left(\mathrm{Tr}\left(\rho[\Gamma] H \right) - \Omega^A \mathrm{Tr}\left(\rho[\Gamma] P_A \right) - \int\d^{d-1}\theta \sqrt{-\sigma}J^I\mathrm{Tr}\left(\rho_t[\Gamma]O_I \right) \right)\delta B.
\end{align}
Hence, \eqref{eq: I variation} and \eqref{eq: Z variation} are contradictory when they are compared via \eqref{eq: Euclidean Hamiltonian expectation} and \eqref{eq: Euclidean momentum expectation}.
Here, \eqref{eq: Euclidean Hamiltonian expectation} and \eqref{eq: Euclidean momentum expectation} come from \eqref{eq: stress and BY}.

Therefore, although source insertion in \eqref{eq: source to metric} seems widely accepted as far as I recognize, we need to include the metric variation to the other source terms if they are not turned off.
That means, \eqref{eq: stress and BY} is modified as
\begin{align}\label{eq: EM modified}
 \mathrm{Tr}\left(\rho_t T_{ab} \right) - \mathrm{Tr}\left(\rho_t O_I(z) \right)w^I(z) \sigma_{ab} = \lim_{R \to \infty} \left(\frac{R}{L} \right)^{d-2}Y_{ab}(z),
\end{align}
which resolves the above contradiction.
In fact, \eqref{eq: EM modified} can also be confirmed by comparing the computations in the Einstein-scalar theory and in the CFT.
One may refer to \cite{in_preparation} where such an example is demonstrated.

Unfortunately, however, the story was not so simple.
For the $\mathrm{U}(1)$ gauge field in the bulk, instead of \eqref{eq: I variation}, we have from \eqref{eq: modified unscaled version}
\begin{align}\label{eq: I variation with gauge}
  \delta_B I_{\mathrm{grav}}^{(E)} = \left(h[\Gamma] - \Omega^A p_A[\Gamma]- \int\d^{d-1}\theta\sqrt{\tilde \sigma}q[\Gamma](\theta) a(\theta) \right) \delta B,
\end{align}
where recall $\sqrt{\tilde \sigma} = \sqrt{-\sigma}$.
This was, as a matter of fact, derived also in \cite{Booth:1999yp} under more general situations.
Now, \eqref{eq: I variation with gauge} is consistent with the form in \eqref{eq: Z variation} without modifying \eqref{eq: stress and BY}.

Therefore, I think that the dictionary between the stress tensor and the Brown-York tensor must be studied more carefully.

Finally, let me explain the interpretation of the examples in section 4, in the light of the above discussion.
Since we do not know the matter theory that makes up the null ray in the Vaidya spacetime, it is not obvious whether the Brown-York tensor follows \eqref{eq: stress and BY} or \eqref{eq: EM modified}.
Thus, it is possible that $h_t$ is not dual to $\mathrm{Tr}(\rho_t H)$, while the masses of the \textit{Euclidean} black holes in section 4 purely measure $\mathrm{Tr}(\rho_\reft H)$ because we chose not to respect the matter field, meaning that the coarse-graining is not properly done.
However, it becomes correct only while the source is turned off, in other words, during the periods when $\mu(v)$ and the other parameters are constant.
Thus, by considering turning on and off the source, our coarse-graining is correct and its associated second law must be satisfied while the source is off.

\subsubsection*{Acknowledgement}
I thank Masafumi Fukuma, Osamu Fukushima, Yuji Hirono, Takanori Ishii, Tomohiro Shigemura, Keito Shimizu, Sotaro Sugishita, Ryota Watanabe, and Takuya Yoda for discussion.
I am also grateful for daily advice by Koji Hashimoto and Shigeki Sugimoto.
My work was supported by Grant-in-Aid for JSPS Fellows No.\ 22KJ1944.

\appendix
\section{Derivation of \eqref{eq: wavefunction of TFD}}\label{app: derivation}
We derive \eqref{eq: wavefunction of TFD} in this appendix.\footnote{
Based on discussion with Osamu Fukushima, Takanori Ishii, Tomohiro Shigemura, Keito Shimizu, Sotaro Sugishita, and Ryota Watanabe.
}
We denote the r.h.s.\ as $T$, and will show that $T$ is deformed to the l.h.s.
In the following, the CPT operator $\Theta$ plays a key role.
The CPT conjugate state $\bra{\Theta\psi}$ of a state $\ket{\psi}$ is defined as
\begin{align}
  \forall \ket{\psi'},\qquad \braket{\Theta \psi | \psi'} = (\braket{\psi' | \Theta | \psi})^*,
\end{align}
where the asterisk means the complex conjugation.
Note that $\Theta$ is antiunitary.

With $\Theta$ and any orthonormal basis $\{\ket{n}\}$, $T$ is deformed as follows:
\begin{align}
  T =& \braket{\phi'| U(t) e^{-\beta \tilde H/2} \Theta^\dagger \Theta V(t) | \phi} =  \bra{\phi'} U(t) e^{-\beta \tilde H/2} \Theta^\dagger \left(\sum_{n}\ketbra{n}{n} \Theta V(t) \ket{\phi} \right)\nonumber\\
  =& \sum_{n} \braket{n|\Theta V(t)|\phi}^* \braket{\phi'|U(t)e^{-\beta \tilde H/2}|n_\Theta}.
\end{align}
Here, we have set $\ket{n_\Theta}:= \Theta^\dagger \ket{n}$.
Assuming that the theory without sources is CPT-invariant, i.e, $H$ in \eqref{eq: tilde H} is invariant, we obtain
\begin{align}
  \braket{n|\Theta V(t)|\phi}^* = \braket{n|V(t)^\dagger \Theta|\phi}^* =  \braket{\Theta \phi| V(t)|n}.
\end{align}
In addition, being a c-number, this can be seen calculated in a copied Hilbert space, which we call as QFT$_{\mathrm{L}}$.
The remaining part is kept in the original Hilbert space, QFT$_{\mathrm{R}}$.
Therefore we conclude
\begin{align}
  T =  {}_{\mathrm{L}}\hspace{-2.5pt} \bra{\Theta \phi}\otimes {}_{\mathrm{R}}\hspace{-2.5pt}\bra{\phi'} V_{\mathrm{L}}(t) \otimes U_{\mathrm{R}}(t) \left(\sum_n \ket n_\mathrm{L}\otimes e^{-\beta \tilde H_\mathrm{R}/2} \ket {n_\Theta}_\mathrm{R} \right).
  \label{eq: T}
\end{align}
In this expression, $\bra{\Theta \phi}$ and $\ket{n_\Theta}$ appear, instead of $\bra{\phi}$ and $\ket{n}$.
However, those differences do not matter.
First, $\phi$ is just a dummy variable in the generating functional \eqref{eq: Lorentz generating functional}.
Second, either $\ket{n}$ or $\ket{n_\Theta}$ can guarantee \eqref{eq: left trace out}, meaning that we could have started from \eqref{eq: TFD} with $\ket{n}_{\mathrm{R}} \to \ket{n_\Theta}_{\mathrm{R}}$ and reach the same results in the main text.
Thus, for simplicity, we have ignored those subtleties in \eqref{eq: wavefunction of TFD}.

\bibliographystyle{jhep} 
\bibliography{ref}

\providecommand{\href}[2]{#2}\begingroup\raggedright\begin{thebibliography}{10}

\bibitem{BardeenCarterHawking}
J.M.~Bardeen, B.~Carter and S.W.~Hawking, \emph{The four laws of black hole
  mechanics}, {\emph{Communications in Mathematical Physics} {\bfseries 31}
  (1973) 161}.

\bibitem{HawkingHartle1st}
S.W.~Hawking and J.B.~Hartle, \emph{Energy and angular momentum flow into a
  black hole}, {\emph{Communications in Mathematical Physics} {\bfseries 27}
  (1972) 283}.

\bibitem{Wald:1993nt}
R.M.~Wald, \emph{{Black hole entropy is the Noether charge}},
  \href{https://doi.org/10.1103/PhysRevD.48.R3427}{\emph{Phys. Rev. D}
  {\bfseries 48} (1993) R3427}
  [\href{https://arxiv.org/abs/gr-qc/9307038}{{\ttfamily gr-qc/9307038}}].

\bibitem{Bekenstein1}
J.D.~Bekenstein, \emph{Black holes and the second law}, {\emph{Lettere al Nuovo
  Cimento (1971-1985)} {\bfseries 4} (1972) 737}.

\bibitem{Jacobson:1995ab}
T.~Jacobson, \emph{{Thermodynamics of space-time: The Einstein equation of
  state}}, \href{https://doi.org/10.1103/PhysRevLett.75.1260}{\emph{Phys. Rev.
  Lett.} {\bfseries 75} (1995) 1260}
  [\href{https://arxiv.org/abs/gr-qc/9504004}{{\ttfamily gr-qc/9504004}}].

\bibitem{HawkingRad}
S.W.~Hawking, \emph{Particle creation by black holes}, {\emph{Communications in
  Mathematical Physics} {\bfseries 43} (1975) 199}.

\bibitem{Strominger:1996sh}
A.~Strominger and C.~Vafa, \emph{{Microscopic origin of the Bekenstein-Hawking
  entropy}}, \href{https://doi.org/10.1016/0370-2693(96)00345-0}{\emph{Phys.
  Lett. B} {\bfseries 379} (1996) 99}
  [\href{https://arxiv.org/abs/hep-th/9601029}{{\ttfamily hep-th/9601029}}].

\bibitem{Hawking2nd}
S.W.~Hawking, \emph{Black holes in general relativity}, {\emph{Communications
  in Mathematical Physics} {\bfseries 25} (1972) 152}.

\bibitem{PhysRevD.7.2333}
J.D.~Bekenstein, \emph{Black holes and entropy},
  \href{https://doi.org/10.1103/PhysRevD.7.2333}{\emph{Phys. Rev. D} {\bfseries
  7} (1973) 2333}.

\bibitem{PhysRevD.9.3292}
J.D.~Bekenstein, \emph{Generalized second law of thermodynamics in black-hole
  physics}, \href{https://doi.org/10.1103/PhysRevD.9.3292}{\emph{Phys. Rev. D}
  {\bfseries 9} (1974) 3292}.

\bibitem{Wall:2009wm}
A.C.~Wall, \emph{{Ten Proofs of the Generalized Second Law}},
  \href{https://doi.org/10.1088/1126-6708/2009/06/021}{\emph{JHEP} {\bfseries
  06} (2009) 021} [\href{https://arxiv.org/abs/0901.3865}{{\ttfamily
  0901.3865}}].

\bibitem{Carlip:2014pma}
S.~Carlip, \emph{{Black Hole Thermodynamics}},
  \href{https://doi.org/10.1142/S0218271814300237}{\emph{Int. J. Mod. Phys. D}
  {\bfseries 23} (2014) 1430023}
  [\href{https://arxiv.org/abs/1410.1486}{{\ttfamily 1410.1486}}].

\bibitem{Wall:2018ydq}
A.C.~Wall, \emph{{A Survey of Black Hole Thermodynamics}},
  \href{https://arxiv.org/abs/1804.10610}{{\ttfamily 1804.10610}}.

\bibitem{Sarkar:2019xfd}
S.~Sarkar, \emph{{Black Hole Thermodynamics: General Relativity and Beyond}},
  \href{https://doi.org/10.1007/s10714-019-2545-y}{\emph{Gen. Rel. Grav.}
  {\bfseries 51} (2019) 63} [\href{https://arxiv.org/abs/1905.04466}{{\ttfamily
  1905.04466}}].

\bibitem{PhysRevD.56.6467}
R.M.~Wald, \emph{``nernst theorem'' and black hole thermodynamics},
  \href{https://doi.org/10.1103/PhysRevD.56.6467}{\emph{Phys. Rev. D}
  {\bfseries 56} (1997) 6467}.

\bibitem{Kehle:2022uvc}
C.~Kehle and R.~Unger, \emph{{Gravitational collapse to extremal black holes
  and the third law of black hole thermodynamics}},
  \href{https://arxiv.org/abs/2211.15742}{{\ttfamily 2211.15742}}.

\bibitem{tHooft:1993dmi}
G.~'t~Hooft, \emph{{Dimensional reduction in quantum gravity}}, {\emph{Conf.
  Proc. C} {\bfseries 930308} (1993) 284}
  [\href{https://arxiv.org/abs/gr-qc/9310026}{{\ttfamily gr-qc/9310026}}].

\bibitem{Susskind:1994vu}
L.~Susskind, \emph{{The World as a hologram}},
  \href{https://doi.org/10.1063/1.531249}{\emph{J. Math. Phys.} {\bfseries 36}
  (1995) 6377} [\href{https://arxiv.org/abs/hep-th/9409089}{{\ttfamily
  hep-th/9409089}}].

\bibitem{Bousso:1999dw}
R.~Bousso, \emph{{The Holographic principle for general backgrounds}},
  \href{https://doi.org/10.1088/0264-9381/17/5/309}{\emph{Class. Quant. Grav.}
  {\bfseries 17} (2000) 997}
  [\href{https://arxiv.org/abs/hep-th/9911002}{{\ttfamily hep-th/9911002}}].

\bibitem{Bousso:2002ju}
R.~Bousso, \emph{{The Holographic principle}},
  \href{https://doi.org/10.1103/RevModPhys.74.825}{\emph{Rev. Mod. Phys.}
  {\bfseries 74} (2002) 825}
  [\href{https://arxiv.org/abs/hep-th/0203101}{{\ttfamily hep-th/0203101}}].

\bibitem{Maldacena:1997re}
J.M.~Maldacena, \emph{{The Large N limit of superconformal field theories and
  supergravity}}, \href{https://doi.org/10.1023/A:1026654312961}{\emph{Adv.
  Theor. Math. Phys.} {\bfseries 2} (1998) 231}
  [\href{https://arxiv.org/abs/hep-th/9711200}{{\ttfamily hep-th/9711200}}].

\bibitem{Witten:1998qj}
E.~Witten, \emph{{Anti-de Sitter space and holography}},
  \href{https://doi.org/10.4310/ATMP.1998.v2.n2.a2}{\emph{Adv. Theor. Math.
  Phys.} {\bfseries 2} (1998) 253}
  [\href{https://arxiv.org/abs/hep-th/9802150}{{\ttfamily hep-th/9802150}}].

\bibitem{Gubser:1998bc}
S.S.~Gubser, I.R.~Klebanov and A.M.~Polyakov, \emph{{Gauge theory correlators
  from noncritical string theory}},
  \href{https://doi.org/10.1016/S0370-2693(98)00377-3}{\emph{Phys. Lett. B}
  {\bfseries 428} (1998) 105}
  [\href{https://arxiv.org/abs/hep-th/9802109}{{\ttfamily hep-th/9802109}}].

\bibitem{Ryu:2006bv}
S.~Ryu and T.~Takayanagi, \emph{{Holographic derivation of entanglement entropy
  from AdS/CFT}},
  \href{https://doi.org/10.1103/PhysRevLett.96.181602}{\emph{Phys. Rev. Lett.}
  {\bfseries 96} (2006) 181602}
  [\href{https://arxiv.org/abs/hep-th/0603001}{{\ttfamily hep-th/0603001}}].

\bibitem{Ryu:2006ef}
S.~Ryu and T.~Takayanagi, \emph{{Aspects of Holographic Entanglement Entropy}},
  \href{https://doi.org/10.1088/1126-6708/2006/08/045}{\emph{JHEP} {\bfseries
  08} (2006) 045} [\href{https://arxiv.org/abs/hep-th/0605073}{{\ttfamily
  hep-th/0605073}}].

\bibitem{Hubeny:2007xt}
V.E.~Hubeny, M.~Rangamani and T.~Takayanagi, \emph{{A Covariant holographic
  entanglement entropy proposal}},
  \href{https://doi.org/10.1088/1126-6708/2007/07/062}{\emph{JHEP} {\bfseries
  07} (2007) 062} [\href{https://arxiv.org/abs/0705.0016}{{\ttfamily
  0705.0016}}].

\bibitem{Wall:2012uf}
A.C.~Wall, \emph{{Maximin Surfaces, and the Strong Subadditivity of the
  Covariant Holographic Entanglement Entropy}},
  \href{https://doi.org/10.1088/0264-9381/31/22/225007}{\emph{Class. Quant.
  Grav.} {\bfseries 31} (2014) 225007}
  [\href{https://arxiv.org/abs/1211.3494}{{\ttfamily 1211.3494}}].

\bibitem{Barrella:2013wja}
T.~Barrella, X.~Dong, S.A.~Hartnoll and V.L.~Martin, \emph{{Holographic
  entanglement beyond classical gravity}},
  \href{https://doi.org/10.1007/JHEP09(2013)109}{\emph{JHEP} {\bfseries 09}
  (2013) 109} [\href{https://arxiv.org/abs/1306.4682}{{\ttfamily 1306.4682}}].

\bibitem{Faulkner:2013ana}
T.~Faulkner, A.~Lewkowycz and J.~Maldacena, \emph{{Quantum corrections to
  holographic entanglement entropy}},
  \href{https://doi.org/10.1007/JHEP11(2013)074}{\emph{JHEP} {\bfseries 11}
  (2013) 074} [\href{https://arxiv.org/abs/1307.2892}{{\ttfamily 1307.2892}}].

\bibitem{Engelhardt:2014gca}
N.~Engelhardt and A.C.~Wall, \emph{{Quantum Extremal Surfaces: Holographic
  Entanglement Entropy beyond the Classical Regime}},
  \href{https://doi.org/10.1007/JHEP01(2015)073}{\emph{JHEP} {\bfseries 01}
  (2015) 073} [\href{https://arxiv.org/abs/1408.3203}{{\ttfamily 1408.3203}}].

\bibitem{Penington:2019npb}
G.~Penington, \emph{{Entanglement Wedge Reconstruction and the Information
  Paradox}}, \href{https://doi.org/10.1007/JHEP09(2020)002}{\emph{JHEP}
  {\bfseries 09} (2020) 002}
  [\href{https://arxiv.org/abs/1905.08255}{{\ttfamily 1905.08255}}].

\bibitem{Almheiri:2019psf}
A.~Almheiri, N.~Engelhardt, D.~Marolf and H.~Maxfield, \emph{{The entropy of
  bulk quantum fields and the entanglement wedge of an evaporating black
  hole}}, \href{https://doi.org/10.1007/JHEP12(2019)063}{\emph{JHEP} {\bfseries
  12} (2019) 063} [\href{https://arxiv.org/abs/1905.08762}{{\ttfamily
  1905.08762}}].

\bibitem{Almheiri:2019hni}
A.~Almheiri, R.~Mahajan, J.~Maldacena and Y.~Zhao, \emph{{The Page curve of
  Hawking radiation from semiclassical geometry}},
  \href{https://doi.org/10.1007/JHEP03(2020)149}{\emph{JHEP} {\bfseries 03}
  (2020) 149} [\href{https://arxiv.org/abs/1908.10996}{{\ttfamily
  1908.10996}}].

\bibitem{Penington:2019kki}
G.~Penington, S.H.~Shenker, D.~Stanford and Z.~Yang, \emph{{Replica wormholes
  and the black hole interior}},
  \href{https://doi.org/10.1007/JHEP03(2022)205}{\emph{JHEP} {\bfseries 03}
  (2022) 205} [\href{https://arxiv.org/abs/1911.11977}{{\ttfamily
  1911.11977}}].

\bibitem{Almheiri:2019qdq}
A.~Almheiri, T.~Hartman, J.~Maldacena, E.~Shaghoulian and A.~Tajdini,
  \emph{{Replica Wormholes and the Entropy of Hawking Radiation}},
  \href{https://doi.org/10.1007/JHEP05(2020)013}{\emph{JHEP} {\bfseries 05}
  (2020) 013} [\href{https://arxiv.org/abs/1911.12333}{{\ttfamily
  1911.12333}}].

\bibitem{Page:1993wv}
D.N.~Page, \emph{{Information in black hole radiation}},
  \href{https://doi.org/10.1103/PhysRevLett.71.3743}{\emph{Phys. Rev. Lett.}
  {\bfseries 71} (1993) 3743}
  [\href{https://arxiv.org/abs/hep-th/9306083}{{\ttfamily hep-th/9306083}}].

\bibitem{Page:2013dx}
D.N.~Page, \emph{{Time Dependence of Hawking Radiation Entropy}},
  \href{https://doi.org/10.1088/1475-7516/2013/09/028}{\emph{JCAP} {\bfseries
  09} (2013) 028} [\href{https://arxiv.org/abs/1301.4995}{{\ttfamily
  1301.4995}}].

\bibitem{Hashimoto:2020cas}
K.~Hashimoto, N.~Iizuka and Y.~Matsuo, \emph{{Islands in Schwarzschild black
  holes}}, \href{https://doi.org/10.1007/JHEP06(2020)085}{\emph{JHEP}
  {\bfseries 06} (2020) 085}
  [\href{https://arxiv.org/abs/2004.05863}{{\ttfamily 2004.05863}}].

\bibitem{Takayanagi:2010wp}
T.~Takayanagi and T.~Ugajin, \emph{{Measuring Black Hole Formations by
  Entanglement Entropy via Coarse-Graining}},
  \href{https://doi.org/10.1007/JHEP11(2010)054}{\emph{JHEP} {\bfseries 11}
  (2010) 054} [\href{https://arxiv.org/abs/1008.3439}{{\ttfamily 1008.3439}}].

\bibitem{Kelly:2013aja}
W.R.~Kelly and A.C.~Wall, \emph{{Coarse-grained entropy and causal holographic
  information in AdS/CFT}},
  \href{https://doi.org/10.1007/JHEP03(2014)118}{\emph{JHEP} {\bfseries 03}
  (2014) 118} [\href{https://arxiv.org/abs/1309.3610}{{\ttfamily 1309.3610}}].

\bibitem{Kelly:2014owa}
W.R.~Kelly, \emph{{Deriving the First Law of Black Hole Thermodynamics without
  Entanglement}}, \href{https://doi.org/10.1007/JHEP10(2014)192}{\emph{JHEP}
  {\bfseries 10} (2014) 192} [\href{https://arxiv.org/abs/1408.3705}{{\ttfamily
  1408.3705}}].

\bibitem{Hubeny:2012wa}
V.E.~Hubeny and M.~Rangamani, \emph{{Causal Holographic Information}},
  \href{https://doi.org/10.1007/JHEP06(2012)114}{\emph{JHEP} {\bfseries 06}
  (2012) 114} [\href{https://arxiv.org/abs/1204.1698}{{\ttfamily 1204.1698}}].

\bibitem{Freivogel:2013zta}
B.~Freivogel and B.~Mosk, \emph{{Properties of Causal Holographic
  Information}}, \href{https://doi.org/10.1007/JHEP09(2013)100}{\emph{JHEP}
  {\bfseries 09} (2013) 100} [\href{https://arxiv.org/abs/1304.7229}{{\ttfamily
  1304.7229}}].

\bibitem{Engelhardt:2017wgc}
N.~Engelhardt and A.C.~Wall, \emph{{No Simple Dual to the Causal Holographic
  Information?}}, \href{https://doi.org/10.1007/JHEP04(2017)134}{\emph{JHEP}
  {\bfseries 04} (2017) 134}
  [\href{https://arxiv.org/abs/1702.01748}{{\ttfamily 1702.01748}}].

\bibitem{Engelhardt:2017aux}
N.~Engelhardt and A.C.~Wall, \emph{{Decoding the Apparent Horizon:
  Coarse-Grained Holographic Entropy}},
  \href{https://doi.org/10.1103/PhysRevLett.121.211301}{\emph{Phys. Rev. Lett.}
  {\bfseries 121} (2018) 211301}
  [\href{https://arxiv.org/abs/1706.02038}{{\ttfamily 1706.02038}}].

\bibitem{Grado-White:2017nhs}
B.~Grado-White and D.~Marolf, \emph{{Marginally Trapped Surfaces and AdS/CFT}},
  \href{https://doi.org/10.1007/JHEP02(2018)049}{\emph{JHEP} {\bfseries 02}
  (2018) 049} [\href{https://arxiv.org/abs/1708.00957}{{\ttfamily
  1708.00957}}].

\bibitem{Engelhardt:2018kcs}
N.~Engelhardt and A.C.~Wall, \emph{{Coarse Graining Holographic Black Holes}},
  \href{https://doi.org/10.1007/JHEP05(2019)160}{\emph{JHEP} {\bfseries 05}
  (2019) 160} [\href{https://arxiv.org/abs/1806.01281}{{\ttfamily
  1806.01281}}].

\bibitem{Chandra:2022fwi}
J.~Chandra and T.~Hartman, \emph{{Coarse graining pure states in AdS/CFT}},
  \href{https://doi.org/10.1007/JHEP10(2023)030}{\emph{JHEP} {\bfseries 10}
  (2023) 030} [\href{https://arxiv.org/abs/2206.03414}{{\ttfamily
  2206.03414}}].

\bibitem{Blanco:2013joa}
D.D.~Blanco, H.~Casini, L.-Y.~Hung and R.C.~Myers, \emph{{Relative Entropy and
  Holography}}, \href{https://doi.org/10.1007/JHEP08(2013)060}{\emph{JHEP}
  {\bfseries 08} (2013) 060} [\href{https://arxiv.org/abs/1305.3182}{{\ttfamily
  1305.3182}}].

\bibitem{Banerjee:2014oaa}
S.~Banerjee, A.~Bhattacharyya, A.~Kaviraj, K.~Sen and A.~Sinha,
  \emph{{Constraining gravity using entanglement in AdS/CFT}},
  \href{https://doi.org/10.1007/JHEP05(2014)029}{\emph{JHEP} {\bfseries 05}
  (2014) 029} [\href{https://arxiv.org/abs/1401.5089}{{\ttfamily 1401.5089}}].

\bibitem{Banerjee:2014ozp}
S.~Banerjee, A.~Kaviraj and A.~Sinha, \emph{{Nonlinear constraints on gravity
  from entanglement}},
  \href{https://doi.org/10.1088/0264-9381/32/6/065006}{\emph{Class. Quant.
  Grav.} {\bfseries 32} (2015) 065006}
  [\href{https://arxiv.org/abs/1405.3743}{{\ttfamily 1405.3743}}].

\bibitem{Lashkari:2014kda}
N.~Lashkari, C.~Rabideau, P.~Sabella-Garnier and M.~Van~Raamsdonk,
  \emph{{Inviolable energy conditions from entanglement inequalities}},
  \href{https://doi.org/10.1007/JHEP06(2015)067}{\emph{JHEP} {\bfseries 06}
  (2015) 067} [\href{https://arxiv.org/abs/1412.3514}{{\ttfamily 1412.3514}}].

\bibitem{jarzynski1997nonequilibrium}
C.~Jarzynski, \emph{Nonequilibrium equality for free energy differences},
  {\emph{Physical Review Letters} {\bfseries 78} (1997) 2690}.

\bibitem{jarzynski2000hamiltonian}
C.~Jarzynski, \emph{Hamiltonian derivation of a detailed fluctuation theorem},
  {\emph{Journal of Statistical Physics} {\bfseries 98} (2000) 77}.

\bibitem{evans1993probability}
D.J.~Evans, E.G.D.~Cohen and G.P.~Morriss, \emph{Probability of second law
  violations in shearing steady states}, {\emph{Physical review letters}
  {\bfseries 71} (1993) 2401}.

\bibitem{kurchan1998fluctuation}
J.~Kurchan, \emph{Fluctuation theorem for stochastic dynamics}, {\emph{Journal
  of Physics A: Mathematical and General} {\bfseries 31} (1998) 3719}.

\bibitem{tasaki2000jarzynski}
H.~Tasaki, \emph{Jarzynski relations for quantum systems and some
  applications}, {\emph{arXiv preprint cond-mat/0009244} (2000) }.

\bibitem{esposito2010entropy}
M.~Esposito, K.~Lindenberg and C.~Van~den Broeck, \emph{Entropy production as
  correlation between system and reservoir}, {\emph{New Journal of Physics}
  {\bfseries 12} (2010) 013013}.

\bibitem{Sagawa:2012eqh}
T.~Sagawa, \emph{{Second Law-Like Inequalities with Quantum Relative Entropy:
  An Introduction}},  \href{https://arxiv.org/abs/1202.0983}{{\ttfamily
  1202.0983}}.

\bibitem{Maldacena:2001kr}
J.M.~Maldacena, \emph{{Eternal black holes in anti-de Sitter}},
  \href{https://doi.org/10.1088/1126-6708/2003/04/021}{\emph{JHEP} {\bfseries
  04} (2003) 021} [\href{https://arxiv.org/abs/hep-th/0106112}{{\ttfamily
  hep-th/0106112}}].

\bibitem{Skenderis:2008dg}
K.~Skenderis and B.C.~van Rees, \emph{{Real-time gauge/gravity duality:
  Prescription, Renormalization and Examples}},
  \href{https://doi.org/10.1088/1126-6708/2009/05/085}{\emph{JHEP} {\bfseries
  05} (2009) 085} [\href{https://arxiv.org/abs/0812.2909}{{\ttfamily
  0812.2909}}].

\bibitem{Balasubramanian:1999re}
V.~Balasubramanian and P.~Kraus, \emph{{A Stress tensor for Anti-de Sitter
  gravity}}, \href{https://doi.org/10.1007/s002200050764}{\emph{Commun. Math.
  Phys.} {\bfseries 208} (1999) 413}
  [\href{https://arxiv.org/abs/hep-th/9902121}{{\ttfamily hep-th/9902121}}].

\bibitem{deHaro:2000vlm}
S.~de~Haro, S.N.~Solodukhin and K.~Skenderis, \emph{{Holographic reconstruction
  of space-time and renormalization in the AdS / CFT correspondence}},
  \href{https://doi.org/10.1007/s002200100381}{\emph{Commun. Math. Phys.}
  {\bfseries 217} (2001) 595}
  [\href{https://arxiv.org/abs/hep-th/0002230}{{\ttfamily hep-th/0002230}}].

\bibitem{cmp/1103922135}
S.W.~Hawking and D.N.~Page, \emph{{Thermodynamics of black holes in anti-de
  Sitter space}}, {\emph{Communications in Mathematical Physics} {\bfseries 87}
  (1982) 577 }.

\bibitem{Fursaev:1995ef}
D.V.~Fursaev and S.N.~Solodukhin, \emph{{On the description of the Riemannian
  geometry in the presence of conical defects}},
  \href{https://doi.org/10.1103/PhysRevD.52.2133}{\emph{Phys. Rev. D}
  {\bfseries 52} (1995) 2133}
  [\href{https://arxiv.org/abs/hep-th/9501127}{{\ttfamily hep-th/9501127}}].

\bibitem{Lewkowycz:2013nqa}
A.~Lewkowycz and J.~Maldacena, \emph{{Generalized gravitational entropy}},
  \href{https://doi.org/10.1007/JHEP08(2013)090}{\emph{JHEP} {\bfseries 08}
  (2013) 090} [\href{https://arxiv.org/abs/1304.4926}{{\ttfamily 1304.4926}}].

\bibitem{Solodukhin:1994yz}
S.N.~Solodukhin, \emph{{The Conical singularity and quantum corrections to
  entropy of black hole}},
  \href{https://doi.org/10.1103/PhysRevD.51.609}{\emph{Phys. Rev. D} {\bfseries
  51} (1995) 609} [\href{https://arxiv.org/abs/hep-th/9407001}{{\ttfamily
  hep-th/9407001}}].

\bibitem{Awad:2007me}
A.M.~Awad, \emph{{First law, counterterms and Kerr-AdS(5) black hole}},
  \href{https://doi.org/10.1142/S0218271809014522}{\emph{Int. J. Mod. Phys. D}
  {\bfseries 18} (2009) 405} [\href{https://arxiv.org/abs/0708.3458}{{\ttfamily
  0708.3458}}].

\bibitem{Faraoni:2020lte}
V.~Faraoni, A.~Giusti and B.H.~Fahim, \emph{{Emmy's letter to Santa Claus (and
  a reply): Vaidya geometries and scalar fields with null gradients}},
  \href{https://arxiv.org/abs/2012.09125}{{\ttfamily 2012.09125}}.

\bibitem{Aniceto:2017gtx}
P.~Aniceto and J.V.~Rocha, \emph{{Dynamical black holes in low-energy string
  theory}}, \href{https://doi.org/10.1007/JHEP05(2017)035}{\emph{JHEP}
  {\bfseries 05} (2017) 035}
  [\href{https://arxiv.org/abs/1703.07414}{{\ttfamily 1703.07414}}].

\bibitem{Hollands:2024vbe}
S.~Hollands, R.M.~Wald and V.G.~Zhang, \emph{{The Entropy of Dynamical Black
  Holes}},  \href{https://arxiv.org/abs/2402.00818}{{\ttfamily 2402.00818}}.

\bibitem{in_preparation}
T.~Shigemura, K.~Shimizu, S.~Sugishita, D.~Takeda and T.~Yoda, \emph{{in
  preparation}}, .

\bibitem{Booth:1999yp}
I.~Booth and R.B.~Mann, \emph{{Static and infalling quasilocal energy of
  charged and naked black holes}},
  \href{https://arxiv.org/abs/gr-qc/9907072}{{\ttfamily gr-qc/9907072}}.

\end{thebibliography}\endgroup
\end{document}